\documentclass[aps,prb,twocolumn,superscriptaddress]{revtex4-1}
\usepackage{graphicx}
\usepackage{bm}
\usepackage{amsmath}
\usepackage{amssymb}
\usepackage{amsfonts}

\usepackage[
  colorlinks=true,
  urlcolor=blue,
  linkcolor=blue,
  citecolor=blue
]{hyperref}

\newcommand{\Dih}{{\mathrm{Dih}}}
\newcommand{\Aut}{{\mathrm{Aut}}}
\newcommand{\Hom}{{\mathrm{Hom}}}
\newcommand{\End}{{\mathrm{End}}}
\newcommand{\Cl}{{\mathrm{Cl}}}
\newcommand{\RR}{\mathbb{R}}
\newcommand{\CC}{\mathbb{C}}
\newcommand{\HH}{\mathbb{H}}
\newcommand{\ZZ}{\mathbb{Z}}

\newcommand{\sC}{\mathcal{C}}
\newcommand{\sR}{\mathcal{R}}
\newcommand{\ptwo}{{\oplus 2}}
\newcommand{\hookdownarrow}{\mathbin{\text{\rotatebox[origin=c]{-90}{$\hookrightarrow$}}}}
\newcommand{\veq}{\mathbin{\text{\rotatebox[origin=c]{-90}{$=$}}}}
\newcommand{\twoheaddownarrow}{\mathbin{\text{\rotatebox[origin=c]{-90}{$\twoheadrightarrow$}}}}

\newcommand{\myspace}{{\;\;\:}}
\def \fg {43pt}
\def \fgb {52pt}
\def \fgx {80pt}
\def \fgw {52pt}

\newcommand{\mytitle}{Classification of Crystalline Topological Insulators and Superconductors with Point Group Symmetries}

\begin{document}

\title{\mytitle}

\author{Eyal Cornfeld}
\affiliation{Raymond and Beverly Sackler School of Physics and Astronomy, Tel-Aviv University, IL-69978 Tel Aviv, Israel}

\author{Adam Chapman}
\affiliation{Department of Computer science, Tel Hai Academic College, IL-12208 Upper Galilee, Israel}

\begin{abstract}
Crystalline topological phases have recently attracted a lot of experimental and theoretical attention. Key advances include the complete elementary band representation analyses of crystalline matter by symmetry indicators and the discovery of higher-order hinge and corner states. However, current classification schemes of such phases are either implicit or limited in scope. We present a new scheme for the explicit classification of crystalline topological insulators and superconductors. These phases are protected by crystallographic point group symmetries and are characterized by bulk topological invariants. The classification paradigm generalizes the Clifford algebra extension process of each Altland-Zirnbauer symmetry class and utilizes algebras which incorporate the point group symmetry. Explicit results for all point group symmetries of three-dimensional crystals are presented as well as for all symmorphic layer groups of two-dimensional crystals. We discuss future extensions for treatment of magnetic crystals and defected or higher-dimensional systems as well as weak and fragile invariants.
\end{abstract}

\maketitle

\section{Introduction}\label{sec:intro}

Over the past decades, the discovery of topological phases such as topological insulators, superconductors, and semimetals, have transformed our understanding of condensed matter physics~\cite{Hasan2010Colloquium,Qi2011Topological,Chiu2016Classification,Armitage2018Weyl}. Study of such phenomena is being extensively used as a new tool for classifying phases of matter which can not be distinguished by broken symmetries. Many topological insulators arise in systems of weakly interacting fermions which feature a bulk gap, and topological superconductors are similarly described by the fermionic quasiparticle excitations of a BCS superconductor.

Key aspects of these topological phases are the symmetries possessed by the material in question. If a gapped material possesses only a charge conservation symmetry, it may only realize an integer quantum Hall effect~\cite{Thouless1982Quantized,Avron1983Homotopy} characterized by a $\ZZ$ topological index. However, in the presence of other non-spatial symmetries, such as time-reversal or particle-hole symmetry, many other topological phases are possible~\cite{Kane2005Topological,Qi2009Time,Schnyder2008Classification}. Examples of such phases include, the $\ZZ_2$ two-dimensional and three-dimensional topological insulators~\cite{Hasan2010Colloquium,Qi2011Topological,Kane2005Quantum,Kane2005Topological,Konig2007Quantum,FuKaneMele2007Topological,Moore2007Topological,Roy2009Topological}, and one-dimensional and two-dimensional topological $p$-wave superconductors~\cite{kitaev2001unpaired,Read2000Paired}. 

Many materials in nature are however also characterized by spatial crystalline symmetries which emanate from their crystallographic structure. Over the past several years, these space group and point group symmetries of crystalline matter have been theoretically predicted to host a large and diverse variety of topological phases~\cite{Mong2010Antiferromagnetic,Fu2011Topological,slager2013space,Chiu2013Classification,Benalcazar2014Classification,Varjas2015Bulk,Shiozaki2015Z2,Cho2015Topological,Yang2015Topological,wang2016hourglass,Ezawa2016Hourglass,Varjas2017Space,Yang2017Topological,Wieder2018Wallpaper,Bouhon2017Global,bouhon2017bulk,Kruthoff2017Topological,kruthoff2017topology}. Many of these proposed topological phases have been measured in various experiments~\cite{hsieh2012topological,dziawa2012topological,tanaka2012experimental,xu2012observation} in materials such as \(\mathrm{PbTe}\),  \(\mathrm{Pb_{1-x}Sn_xTe}\), and \(\mathrm{Pb_{1-x}Sn_xSe}\). Other crystalline topological phases host exotic surface behaviour such as hinge and corner states~\cite{parameswaran2017topological,Benalcazar2017Quantized,Benalcazar2017Electric,Song2017d,Langbehn2017Reflection,Schindler2018Higher,schindler2018bismuth,xu2017topological,Shapourian2018Topological,lin2017topological,Ezawa2018Higher,Khalaf2018Higher,Geier2018Second,trifunovic2018higher,fang2017rotation,okuma2018topological}. Such higher-order topological insulators and superconductors have been also recently experimentally observed in bismuth by Schindler \textit{et al.}~[\onlinecite{schindler2018bismuth}].
A recent survey by Vergniory \textit{et al.}~[\onlinecite{vergniory2018high}] had found that a staggering 24\% of materials in nature
have some nontrivial band structure topology.

In this paper we engage the vital challenge of achieving a complete classification of all possible topological phases in presence of all possible material symmetries.

A major milestone in our understanding of topological phases came with the discovery of the periodic table of topological insulators and superconductors~\cite{kitaev2009periodic,schnyder2009classification,ryu2010topological,Hasan2010Colloquium,moore2010birth,stone2010symmetries,Teo2010Topological,franz2013topological,witten2015three,Chiu2016Classification}; see Table~\ref{tab:per}. The table provides a complete classification of all topological phases of non-interacting fermions in presence of non-spatial symmetries. The presence and absence of these symmetries are categorized into the 10 Altland-Zirnbauer (AZ) symmetry classes~\cite{Altland1997Nonstandard,heinzner2005symmetry,zirnbauer2010symmetry} and form the ``tenfold way". Each AZ symmetry class corresponds to a topological ``classifying space" of possible free-fermion Hamiltonians respecting the symmetry, and these are the topological invariants of these classifying spaces which are the $\ZZ$ and $\ZZ_2$ topological indices.

There are many different view-points for this profound classification~\cite{Schnyder2008Classification,schnyder2009classification,ryu2010topological,kitaev2009periodic,stone2010symmetries}. The approach of Refs.~[\onlinecite{kitaev2009periodic,stone2010symmetries,abramovici2012clifford}], which was pioneered by Kitaev~[\onlinecite{kitaev2009periodic}], takes an algebraic perspective. In this paradigm, the set of non-spatial symmetry transformations forms a Clifford algebra with a specific action over the Hilbert space. The enumeration of all possible topologically distinct Hamiltonians is equivalent to asking the following: How many actions can a Hamiltonian have on the Hilbert space which are compatible with the algebra of non-spatial symmetries?
The answer to this question is exactly the ``classifying space" for the AZ symmetry class; see Fig.~\ref{fig:diag1}.
The twofold and eightfold periodic structures within the table then naturally follow from the Bott periodicity of Clifford modules~\cite{atiyah1964clifford}.

\begin{figure}[t]
\centering
\begin{tabular}{ccccccc}
& \begin{tabular}{@{}c@{}} Non-spatial \\ symmetries\end{tabular} &  & \begin{tabular}{@{}c@{}} Symmetries \\ + \\ Hamiltonian\end{tabular} & & & \begin{tabular}{@{}c@{}} Classifying \\ space\end{tabular}\\
& $\Downarrow$ & & $\Downarrow$ & & & $\Uparrow$ \\
$\Bigg($ & \begin{tabular}{@{}c@{}} Clifford \\ algebra\end{tabular} & $\hookrightarrow$ & \begin{tabular}{@{}c@{}}Extended \\ Clifford algebra\end{tabular} & $\Bigg)$ & $\mapsto$ & \begin{tabular}{@{}c@{}} All Hilbert- \\ space actions\end{tabular}
\\
\\
 & $\Cl_{p,q}$ &  & $\Cl_{p+1,q}$ &  &  & $\sR_{p-q+2}$
\end{tabular}
\caption{The Clifford algebra classification scheme for non-spatial symmetries~\cite{kitaev2009periodic,stone2010symmetries,abramovici2012clifford}.}\label{fig:diag1}
\end{figure}

In this paper we generalize this classification scheme as to include the effects of all point group symmetries of crystalline matter.

One highly successful approach towards solving the classification problem is the use of symmetry indicators of elementary band representations~\cite{Dong2016Classification,po2017symmetry,bradlyn2017topological,Watanabe2018Structure,Bradlyn2018Band,song2018quantitative,Cano2018Building,Vergniory2017Graph,Ono2018Unified,vergniory2018high,Khalaf2018Symmetry,Khalaf2018Higher}. In this approach, the ``symmetry indicators", which are a generalization of the Fu-Kane~\cite{Fu2007Topological,Fu2011Topological} formula $\ZZ_2$ invariant, are used to analyse the band structure into elementary band representation of the corresponding crystallographic group. It is shown that bands which form one subpart of a
disconnected elementary band representation must always be topological~\cite{Cano2018Building}. This is a very prolific classification scheme which is used to categorize numerous topological crystalline phases. However, this technique requires one to perform Berry-phase analyses of individual electronic Bloch wavefunctions along various Wilson loops in order to detect different topological phases which correspond to the same band representation, e.g., the integer quantum Hall effect.

An alternate theoretically complete and rigorous approach is the formulation of the crystalline classification problem using twisted equivariant K-theory~\cite{freed2013twisted,Morimoto2013Topological,Shiozaki2014Topology,Hsieh2014CPT,Shiozaki2016Topology,Shiozaki2017Topological,Trifunovic2017Bott,shiozaki2018atiyah,Geier2018Second,trifunovic2018higher,okuma2018topological}. Within twisted equivariant K-theory the crystallographic group action on the Hamiltonian is encoded by a twist on the Brillouin zone (BZ) which serves as the base space of the K-group~\cite{Shiozaki2017Topological}. This approach was successful in obtaining a complete classification of all topological invariants of order-two magnetic space group crystals~\cite{Morimoto2013Topological,Shiozaki2014Topology} as well as of all wallpaper group crystals~\cite{Shiozaki2017Topological,graph}. However, due to the mathematically challenging nature of the paradigm, ongoing works~\cite{shiozaki2018atiyah} are still being carried in an attempt to calculate all topological invariants of all crystallographic symmetry groups.

Herein, we follow the approach of Freed and Moor~[\onlinecite{freed2013twisted}], Morimoto and Furusaki~[\onlinecite{Morimoto2013Topological}], and Shiozaki, Sato, and Gomi~[\onlinecite{Shiozaki2014Topology,Shiozaki2017Topological}]. We focus on bulk topological invariants which are akin to the strong topological invariants of topological insulators and superconductors with non-spatial symmetries~\cite{kitaev2009periodic,schnyder2009classification,ryu2010topological,franz2013topological,witten2015three,abramovici2012clifford}. We generalize the Clifford algebra classification scheme (see Fig.~\ref{fig:diag1}) by incorporating the point group symmetry actions into the algebra of symmetries generated by the non-spatial symmetries, and a Hamiltonian compatible with this symmetry creates a natural $\ZZ_2$-graded algebraic structure. Since any operator, such as the Hamiltonian, can be thought of as a transformation acting on the Hilbert space, the study of all compatible Hamiltonians is thus equivalent to the study of all possible actions of the extended $\ZZ_2$-graded algebra on the Hilbert space; see Fig.~\ref{fig:diag2}.

For each of the possible 32 crystallographic point group symmetries, we find the appropriate $\ZZ_2$-graded algebra and its corresponding ``classifying space" of possible actions. The topological indices characterizing possible topological insulators and superconductors in any of  the 10 AZ symmetry classes are the topological invariants of this classifying space, and these invariants are herein presented.

The majority of phases found by our classification paradigm correspond to novel crystalline topological insulators and superconductors. Nevertheless, a tremendous amount of work on various crystalline topological phases has recently been carried out by numerous research groups around the globe, we thus also compare our results to many phases that had already been analysed (see Sec.~\ref{sec:discussion}).

The rest of the paper is divided as follows: In Sec.~\ref{sec:results} we summarize our results which are presented in the tables throughout the paper. In Sec.~\ref{sec:meth} we present the physical paradigm of our classification and then provide the mathematical background needed for the analysis. In Sec.~\ref{sec:exam} we give some pedagogical examples for calculating the topological classification of all symmetries. In Sec.~\ref{sec:discussion} we discuss our results in comparison with previous classification schemes and suggest possible future extensions.

\begin{figure}[t]
\centering
\begin{tabular}{ccccccc}
& \begin{tabular}{@{}c@{}} Crystalline \\  \& non-spatial \\ symmetries\end{tabular} &  & \begin{tabular}{@{}c@{}} Symmetries \\ + \\ Hamiltonian\end{tabular} & & & \begin{tabular}{@{}c@{}} Classifying \\ space\end{tabular}\\
& $\Downarrow$ & & $\Downarrow$ & & & $\Uparrow$ \\
$\Bigg($ & \begin{tabular}{@{}c@{}} Algebra of \\ symmetries\end{tabular} & $\hookrightarrow$ & \begin{tabular}{@{}c@{}} $\ZZ_2$-graded \\ real algebra\end{tabular} & $\Bigg)$ & $\mapsto$ & \begin{tabular}{@{}c@{}} All Hilbert- \\ space actions\end{tabular}
\\
\\
 & $B^0$ &  & $B$ & & & 
\end{tabular}
\caption{Our $\ZZ_2$-graded algebra classification scheme for crystalline symmetries.}\label{fig:diag2}
\end{figure}

\section{Summary of Results}\label{sec:results}

\begin{table*}[t]
	\begin{equation*}
	\begin{array}{l||c||l||l||c|c|c|c}
	q & \mathrm{Extension}						& \mathrm{Classifying~space}				& \mathrm{AZ~class}	& d=0	& d=1	& d=2	& d=3
	\\ \hline\hline
	0 & \CC\hookrightarrow\CC\oplus\CC			& \sC_0=\prod_{k+m=n} U(n)/(U(k)\times U(m))	& \mathrm{A} 	& \ZZ	& 0		& \ZZ	& 0								
	\\ \hline
	1 & \CC\oplus\CC\hookrightarrow M_2(\CC)	& \sC_1=U(n)									& \mathrm{AIII} & 0		& \ZZ	& 0		& \ZZ									
	\\ \hline\hline
	0 & \RR\hookrightarrow\RR\oplus\RR			& \sR_0=\prod_{k+m=n} O(n)/(O(k)\times O(m))	& \mathrm{AI} 	& \ZZ	& 0 	& 0		& 0						
	\\ \hline
	1 & \RR\oplus\RR\hookrightarrow M_2(\RR)	& \sR_1=O(n)									& \mathrm{BDI} 	& \ZZ_2	& \ZZ	& 0		& 0										
	\\ \hline
	2 & \RR\hookrightarrow\CC					& \sR_2=O(2n)/U(n)								& \mathrm{D} 	& \ZZ_2	& \ZZ_2	& \ZZ	& 0							
	\\ \hline
	3 & \CC\hookrightarrow\HH					& \sR_3=U(2n)/Sp(n)								& \mathrm{DIII} & 0		& \ZZ_2	& \ZZ_2	& \ZZ							
	\\ \hline
	4 & \HH\hookrightarrow\HH\oplus\HH			& \sR_4=\prod_{k+m=n} Sp(n)/(Sp(k)\times Sp(m))	& \mathrm{AII} 	& \ZZ	& 0		& \ZZ_2	& \ZZ_2	
	\\ \hline
	5 & \HH\oplus\HH\hookrightarrow M_2(\HH)	& \sR_5=Sp(n)									& \mathrm{CII} 	& 0		& \ZZ	& 0		& \ZZ_2										
	\\ \hline
	6 & \HH\hookrightarrow M_2(\CC)				& \sR_6=Sp(n)/U(n)								& \mathrm{C} 	& 0		& 0		& \ZZ	& 0						
	\\ \hline
	7 & \CC\hookrightarrow M_2(\RR)				& \sR_7=U(n)/O(n)								& \mathrm{CI} 	& 0		& 0		& 0		& \ZZ											
	\end{array}
	\end{equation*}
	\caption{Bulk topological invariants, \(\pi_0(\sR_{q-d})\) and \(\pi_0(\sC_{q-d})\), for real and complex Altland-Zirnbauer symmetry classes, \(q\), in \(d\) spatial dimensions.}\label{tab:per}
\end{table*}

The results are brought for all point group symmetries in all crystal systems, triclinic, monoclinic, orthorhombic, tetragonal, trigonal, hexagonal, and cubic.
All groups are brought both in Sch\"{o}nflies (Sch\"{o}n.) notation and in Hermann–Mauguin (HM) notation, they are accompanied by a solid displaying the crystalline symmetry group, taken with permission from Ashcroft and Mermin~\cite{ashcroft1976solid}. Bulk topological invariants for all AZ symmetry classes are presented in Tables~\ref{tab:p1}-\ref{tab:m1}. All classifying spaces for all three-dimensional (3D) point group symmetries in all crystal systems are compactly presented in Table~\ref{tab:class}. In Appendix~\ref{app:layer} we use our techniques to calculate all classifying spaces for all two-dimensional (2D) symmorphic layer group symmetries~\cite{graph}; these results are also compactly presented in Table~\ref{tab:class}. Possible extensions of our work in treatment magnetic crystals as well as defected and higher-dimensional systems are discussed in Sec.~\ref{sec:discussion}.

\section{Methodology of Classification}\label{sec:meth}

\subsection{Physical Paradigm}\label{sec:paradigm}


The many-body Hilbert space is a complex linear vector space. A symmetry action on the states within the Hilbert space is either a unitary or anti-unitary action. These imply that the Hilbert space forms a real module over the algebra of symmetries, denoted \(B^0\). When no crystalline symmetry is present, \(B^0\) is generated by some selection of the following non-spatial symmetries~\cite{schnyder2009classification,ryu2010topological,kennedy2016bott}: charge conservation \(Q\), time reversal \(T\), particle-hole symmetry \(C\), and spin rotations \(S_1,S_2,S_3\). Such a selection corresponds to an AZ symmetry class~\cite{Altland1997Nonstandard,heinzner2005symmetry,zirnbauer2010symmetry}; see Table~\ref{tab:per}. Any translation invariant non-interacting quantum dynamics is described by a Hamiltonian, \(\mathcal{H}(\mathbf{k})\), which is quadratic in creation-annihilation(/Majorana) operators of the many-body Fock(/Nambu) space representation. When using this representation, the Hamiltonian may be linearized around the $\Gamma$-point, which is the high-symmetry time-reversal-invariant point, such that~\cite{kitaev2009periodic,schnyder2009classification,ryu2010topological,franz2013topological,witten2015three,abramovici2012clifford}
\begin{equation}\label{gammapt}
\mathcal{H}(\mathbf{k})=iM+\bm{\gamma}\cdot\mathbf{k},\myspace\{\gamma_i,\gamma_j\}=2\delta_{ij}.
\end{equation}
Here, the Dirac gamma matrices \(\bm{\gamma}=\gamma_1,\gamma_2,\ldots,\gamma_d\) (in \(d\) spatial dimensions) may be naturally incorporated into the algebra of symmetries, \(B^0\).
When studying either insulators or superconductors which correspond to gapped Hamiltonians, \(\mathcal{H}(\mathbf{k})\), the mass matrix, $M$, which gaps the spectrum, may be spectrally flattened (i.e. ${M^2=-1}$), and, together with the algebra of symmetries, \(B^0\), form an extended algebra, \(B\). Classifying all gapped Dirac Hamiltonians is thus equivalent to classifying all Hilbert space actions~\cite{abramovici2012clifford} of the extended algebra \(B\) which are compatible with the action of \(B^0\); see Figs.~\ref{fig:diag1} and \ref{fig:diag2}.
One finds a Morita equivalent Clifford algebra structure~\cite{kitaev2009periodic,atiyah1964clifford,Vela2002Central,cliff} for both algebras, \({B^0=\Cl_{p,q}}\) and \({B=\Cl_{p+1,q}}\). Moreover, one may also identify the extended algebra as a \(\ZZ_2\)-graded algebra (superalgebra) \(B=B^0\oplus B^1\) whose even part is the algebra of symmetries \(B^0\); see Sec.~\ref{sec:graded}. The space of Dirac Hamiltonians corresponding to the extension \({\Cl_{q+6,d}\hookrightarrow\Cl_{q+7,d}}\) is stably homotopic to a Cartan symmetric space \(\sR_{q-d}\) which is the classifying space~\cite{kitaev2009periodic,schnyder2009classification,ryu2010topological,Hasan2010Colloquium,moore2010birth,franz2013topological,witten2015three} of the AZ symmetry class \(q\) in \(d\) dimensions. Every stable topological phase is therefore characterized by an invariant in the group \(\pi_0(\sR_{q-d})\); see Sec.~\ref{sec:ext}. This structure is summarized in the periodic table~\cite{kitaev2009periodic,schnyder2009classification,ryu2010topological,Hasan2010Colloquium,moore2010birth,franz2013topological,witten2015three} of topological insulators and superconductors, which follows from the Atiyah-Bott-Shapiro construction~\cite{kitaev2009periodic,atiyah1964clifford}; see Table~\ref{tab:per}. This classification of Dirac Hamiltonians captures the bulk topological invariants corresponding to strong topological phases~\cite{kitaev2009periodic,schnyder2009classification,ryu2010topological,franz2013topological,witten2015three,abramovici2012clifford}, and it is these bulk strong topological invariants we herein generalize to incorporate spatial crystalline symmetry. Possible future analyses of weak~\cite{FuKaneMele2007Topological,Moore2007Topological,Varjas2017Space} and fragile~\cite{Po2018Fragile,bouhon2018wilson,bradlyn2018disconnected} crystalline topological phases are discussed in Sec.~\ref{sec:discussion}.

\begin{table*}[t]
	\begin{equation*}
	\begin{array}{l||c|c||c|c|c||c|c|c||}
	&
	\parbox{\fgw}{\includegraphics[width=\fgw]{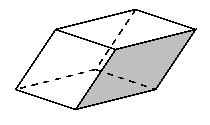}} & 
	\parbox{\fgw}{\includegraphics[width=\fgw]{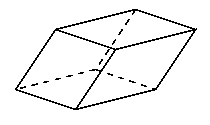}} & 
	\parbox{\fgw}{\includegraphics[width=\fgw]{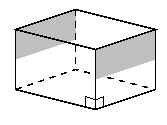}} & 
	\parbox{\fgw}{\includegraphics[width=\fgw]{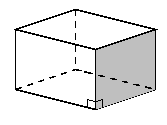}} & \parbox{\fgw}{\includegraphics[width=\fgw]{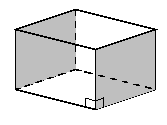}} & \parbox{\fgw}{\includegraphics[height=\fgb]{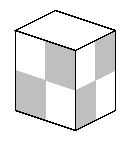}} & \parbox{\fgw}{\includegraphics[height=\fgb]{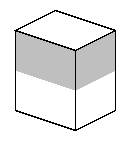}} & \parbox{\fgw}{\includegraphics[height=\fgb]{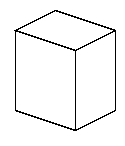}}
	\\ \hline \hline
	\mathrm{Sch\ddot{o}n.}	& C_1 & C_i,S_2 & C_2,D_1 & C_s,C_{1h},C_{1v} & C_{2h},D_{1d} & D_2 & C_{2v},D_{1h} & D_{2h}
	\\ \hline
	\mathrm{HM}	& 1 & \bar 1 & 2 & \mathrm{m} & 2/\mathrm{m} & 222 & \mathrm{mm}2 & \mathrm{mmm}
	\\ \hline\hline
	& \sC_{q+1}	& \sC_{q} & \sC_{q+1}^2 & \sC_{q} & \sC_{q}^2 & \sC_{q+1}^4 & \sC_{q}^2	& \sC_{q}^4
	\\ \hline\hline
	\mathrm{A} 		& 0		& \ZZ	& 0			& \ZZ	& \ZZ^2		& 0			& \ZZ^2		& \ZZ^4							
	\\ \hline
	\mathrm{AIII} 	& \ZZ	& 0		& \ZZ^2		& 0		& 0			& \ZZ^4		& 0			& 0									
	\\ \hline\hline
	& \sR_{q-3}	& \sR_{q-4} & \sR_{q-3}^2 & \sR_{q-4} & \sR_{q-4}^2 & \sR_{q-3}^4 & \sR_{q-4}^2	& \sR_{q-4}^4
	\\ \hline\hline
	\mathrm{AI} 	& 0		& \ZZ	& 0			& \ZZ	& \ZZ^2		& 0			& \ZZ^2		& \ZZ^4						
	\\ \hline
	\mathrm{BDI} 	& 0		& 0		& 0			& 0		& 0			& 0			& 0			& 0									
	\\ \hline
	\mathrm{D} 		& 0		& 0		& 0			& 0		& 0			& 0			& 0			& 0						
	\\ \hline
	\mathrm{DIII} 	& \ZZ	& 0		& \ZZ^2		& 0		& 0			& \ZZ^4		& 0			& 0						
	\\ \hline
	\mathrm{AII} 	& \ZZ_2	& \ZZ	& \ZZ_2^2	& \ZZ	& \ZZ^2		& \ZZ_2^4	& \ZZ^2		& \ZZ^4	
	\\ \hline
	\mathrm{CII} 	& \ZZ_2	& \ZZ_2	& \ZZ_2^2	& \ZZ_2	& \ZZ_2^2	& \ZZ_2^4	& \ZZ_2^2	& \ZZ_2^4										
	\\ \hline
	\mathrm{C} 		& 0		& \ZZ_2	& 0			& \ZZ_2	& \ZZ_2^2	& 0			& \ZZ_2^2	& \ZZ_2^4						
	\\ \hline
	\mathrm{CI} 	& \ZZ	& 0		& \ZZ^2		& 0		& 0			& \ZZ^4		& 0			& 0										
	\end{array}
	\end{equation*}
	\caption{Bulk topological invariants and classifying spaces for all AZ symmetry classes with the following symmetries of crystalline matter: the triclinic crystal system point group symmetries \(C_1,C_i\), the monoclinic crystal system point group symmetries \(C_2,C_s,C_{2h}\), and the orthorhombic crystal system point group symmetries \(D_2,C_{2v},D_{2h}\).}\label{tab:p1}\label{tab:o2}
\end{table*}

\begin{table*}[t]
	\begin{equation*}
	\begin{array}{l||c||c|c|c|c|c|c|c||}
	\multicolumn{2}{l||}{} &
	\parbox{\fgw}{\includegraphics[height=\fgb]{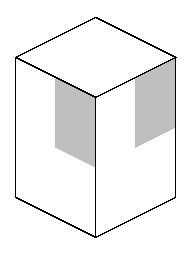}} & 
	\parbox{\fgw}{\includegraphics[height=\fgb]{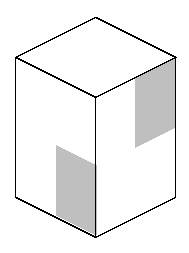}} & 
	\parbox{\fgw}{\includegraphics[height=\fgb]{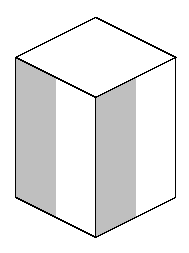}} & 
	\parbox{\fgw}{\includegraphics[height=\fgb]{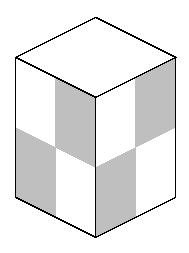}} & 
	\parbox{\fgw}{\includegraphics[height=\fgb]{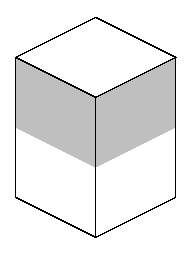}} & 
	\parbox{\fgw}{\includegraphics[height=\fgb]{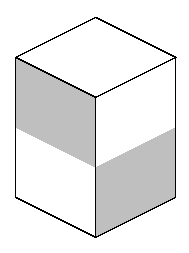}} & 
	\parbox{\fgw}{\includegraphics[height=\fgb]{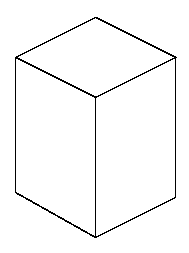}}
	\\ \hline\hline
	\mathrm{Sch\ddot{o}n.}	& C_1 & C_4 & S_4 & C_{4h} & D_4 & C_{4v} & D_{2d} & D_{4h}
	\\ \hline
	\mathrm{HM}	& 1 & 4 & \bar 4 & 4/\mathrm{m} & 422 & 4\mathrm{mm} & \bar{4}2\mathrm{m} & 4/\mathrm{mmm}
	\\ \hline\hline
	& \sC_{q+1}	& \sC_{q+1}^4 & \sC_{q}^2 & \sC_q^4 & \sC_{q+1}^5 & \sC_{q}^2\times\sC_{q+1} & \sC_{q}^2\times\sC_{q+1}	& \sC_{q}^5
	\\ \hline\hline
	\mathrm{A} 		& 0		& 0					& \ZZ^2				& \ZZ^4				& 0			& \ZZ^2				& \ZZ^2				& \ZZ^5
	\\ \hline
	\mathrm{AIII} 	& \ZZ	& \ZZ^4				& 0					& 0					& \ZZ^5		& \ZZ				& \ZZ				& 0
	\\ \hline\hline
	& \sR_{q-3}	& \sR_{q-3}^2\times\sC_{q+1} & \sR_{q-4}\times\sR_{q-2} & \sR_{q-4}^2\times\sC_q & \sR_{q-3}^5 & \sR_{q-4}^2\times\sR_{q-5} & \sR_{q-4}^2\times\sR_{q-3}	& \sR_{q-4}^5
	\\ \hline\hline
	\mathrm{AI} 	& 0		& 0					& \ZZ				& \ZZ^3				& 0			& \ZZ^2				& \ZZ^2				& \ZZ^5
	\\ \hline
	\mathrm{BDI} 	& 0		& \ZZ				& 0					& 0					& 0			& \ZZ				& 0					& 0
	\\ \hline
	\mathrm{D} 		& 0		& 0					& \ZZ				& \ZZ				& 0			& 0					& 0					& 0
	\\ \hline
	\mathrm{DIII} 	& \ZZ	& \ZZ^3				& \ZZ_2				& 0					& \ZZ^5		& 0					& \ZZ				& 0
	\\ \hline
	\mathrm{AII} 	& \ZZ_2	& \ZZ_2^2			& \ZZ\times\ZZ_2	& \ZZ^3				& \ZZ_2^5	& \ZZ^2				& \ZZ^2\times\ZZ_2	& \ZZ^5 
	\\ \hline
	\mathrm{CII} 	& \ZZ_2	& \ZZ_2^2\times\ZZ	& \ZZ_2				& \ZZ_2^2			& \ZZ_2^5	& \ZZ_2^2\times\ZZ	& \ZZ_2^3			& \ZZ_2^5		
	\\ \hline
	\mathrm{C} 		& 0		& 0					& \ZZ_2\times\ZZ	& \ZZ_2^2\times\ZZ	& 0			& \ZZ_2^3			& \ZZ_2^2			& \ZZ_2^5
	\\ \hline
	\mathrm{CI} 	& \ZZ	& \ZZ^3				& 0				& 0					& \ZZ^5		& \ZZ_2				& \ZZ				& 0	
	\end{array}
	\end{equation*}
	\caption{Bulk topological invariants and classifying spaces for all AZ symmetry classes with all tetragonal crystal system point group symmetries of crystalline matter,  \(C_4,S_4,C_{4h},D_4,C_{4v},D_{2d},D_{4h}\). }\label{tab:exs4}
\end{table*}

\begin{table*}[t]
	\begin{equation*}
	\begin{array}{l||c||c|c|c|c|c||}
	\multicolumn{2}{l||}{} &
	\parbox{\fgw}{\includegraphics[height=\fgb]{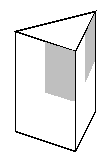}} & 
	\parbox{\fgw}{\includegraphics[height=\fgb]{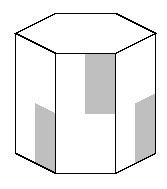}} & 
	\parbox{\fgw}{\includegraphics[height=\fgb]{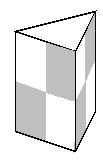}} & 
	\parbox{\fgw}{\includegraphics[height=\fgb]{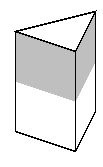}} & 
	\parbox{\fgw}{\includegraphics[height=\fgb]{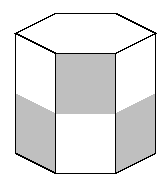}}
	\\ \hline\hline
	\mathrm{Sch\ddot{o}n.}	& C_1 & C_3 & C_{3i},S_6 & D_3 & C_{3v} & D_{3d}
	\\ \hline
	\mathrm{HM}	& 1 & 3 & \bar 3 & 32 & 3\mathrm{m} & \bar{3}\mathrm{m}
	\\ \hline\hline
	& \sC_{q+1}	& \sC_{q+1}^3 & \sC_{q}^3 & \sC_{q+1}^3 & \sC_{q}\times\sC_{q+1} & \sC_{q}^3
	\\ \hline\hline
	\mathrm{A} 		& 0		& 0					& \ZZ^3				& 0			& \ZZ				& \ZZ^3
	\\ \hline
	\mathrm{AIII} 	& \ZZ	& \ZZ^3				& 0					& \ZZ^3		& \ZZ				& 0
	\\ \hline\hline
	& \sR_{q-3}	& \sR_{q-3}\times\sC_{q+1} & \sR_{q-4}\times\sC_{q} & \sR_{q-3}^3 & \sR_{q-4}\times\sR_{q-5} & \sR_{q-4}^3
	\\ \hline\hline
	\mathrm{AI} 	& 0		& 0					& \ZZ^2				& 0			& \ZZ				& \ZZ^3
	\\ \hline
	\mathrm{BDI} 	& 0		& \ZZ				& 0					& 0			& \ZZ				& 0
	\\ \hline
	\mathrm{D} 		& 0		& 0					& \ZZ				& 0			& 0					& 0
	\\ \hline
	\mathrm{DIII} 	& \ZZ	& \ZZ^2				& 0					& \ZZ^3		& 0					& 0
	\\ \hline
	\mathrm{AII} 	& \ZZ_2	& \ZZ_2				& \ZZ^2				& \ZZ_2^3	& \ZZ				& \ZZ^3 
	\\ \hline
	\mathrm{CII} 	& \ZZ_2	& \ZZ_2\times\ZZ	& \ZZ_2				& \ZZ_2^3	& \ZZ_2\times\ZZ	& \ZZ_2^3		
	\\ \hline
	\mathrm{C} 		& 0		& 0					& \ZZ_2\times\ZZ	& 0			& \ZZ_2^2			& \ZZ_2^3
	\\ \hline
	\mathrm{CI} 	& \ZZ	& \ZZ^2				& 0					& \ZZ^3		& \ZZ_2				& 0	
	\end{array}
	\end{equation*}
	\caption{Bulk topological invariants and classifying spaces for all AZ symmetry classes with all trigonal crystal system point group symmetries of crystalline matter, \(C_3,C_{3i},D_3,C_{3v},D_{3d}\). }\label{tab:exc3}
\end{table*}

\begin{table*}[t]
	\begin{equation*}
	\begin{array}{l||c||c|c|c|c|c|c|c||}
	\multicolumn{2}{l||}{} &
	\parbox{\fgw}{\includegraphics[height=\fgb]{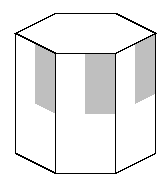}} & 
	\parbox{\fgw}{\includegraphics[height=\fgb]{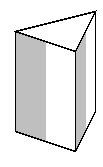}} & 
	\parbox{\fgw}{\includegraphics[height=\fgb]{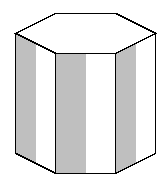}} & 
	\parbox{\fgw}{\includegraphics[height=\fgb]{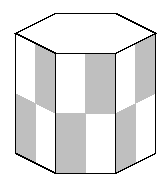}} & 
	\parbox{\fgw}{\includegraphics[height=\fgb]{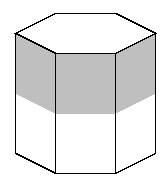}} & 
	\parbox{\fgw}{\includegraphics[height=\fgb]{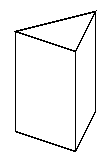}} & 
	\parbox{\fgw}{\includegraphics[height=\fgb]{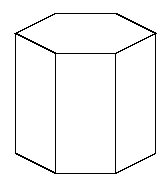}}
	\\ \hline\hline
	\mathrm{Sch\ddot{o}n.}	& C_1 & C_6 & C_{3h} & C_{6h} & D_6 & C_{6v} & D_{3h} & D_{6h}
	\\ \hline
	\mathrm{HM}	& 1 & 6 & 3/\mathrm{m} & 6/\mathrm{m} & 622 & 6\mathrm{mm} & \bar{6}\mathrm{m}2 & 6/\mathrm{mmm}
	\\ \hline\hline
	& \sC_{q+1}	& \sC_{q+1}^6 & \sC_{q}^3 & \sC_q^6 & \sC_{q+1}^6 & \sC_{q}^2\times\sC_{q+1}^2 & \sC_{q}^3	& \sC_{q}^6
	\\ \hline\hline
	\mathrm{A} 		& 0		& 0					& \ZZ^3				& \ZZ^6				& 0			& \ZZ^2				& \ZZ^3				& \ZZ^6
	\\ \hline
	\mathrm{AIII} 	& \ZZ	& \ZZ^6				& 0					& 0					& \ZZ^6		& \ZZ^2				& 0					& 0
	\\ \hline\hline
	& \sR_{q-3}	& \sR_{q-3}^2\times\sC_{q+1}^2 & \sR_{q-4}\times\sC_{q} & \sR_{q-4}^2\times\sC_q^2 & \sR_{q-3}^6 & \sR_{q-4}^2\times\sR_{q-5}^2 & \sR_{q-4}^3	& \sR_{q-4}^6
	\\ \hline\hline
	\mathrm{AI} 	& 0		& 0					& \ZZ^2				& \ZZ^4				& 0			& \ZZ^2				& \ZZ^3				& \ZZ^6
	\\ \hline
	\mathrm{BDI} 	& 0		& \ZZ^2				& 0					& 0					& 0			& \ZZ^2				& 0					& 0
	\\ \hline
	\mathrm{D} 		& 0		& 0					& \ZZ				& \ZZ^2				& 0			& 0					& 0					& 0
	\\ \hline
	\mathrm{DIII} 	& \ZZ	& \ZZ^4				& 0					& 0					& \ZZ^6		& 0					& 0				& 0
	\\ \hline
	\mathrm{AII} 	& \ZZ_2	& \ZZ_2^2			& \ZZ^2				& \ZZ^4				& \ZZ_2^6	& \ZZ^2				& \ZZ^3				& \ZZ^6 
	\\ \hline
	\mathrm{CII} 	& \ZZ_2	&\ZZ_2^2\times\ZZ^2 & \ZZ_2				& \ZZ_2^2			& \ZZ_2^6	&\ZZ_2^2\times\ZZ^2	& \ZZ_2^3				& \ZZ_2^6		
	\\ \hline
	\mathrm{C} 		& 0		& 0					& \ZZ_2\times\ZZ	&\ZZ_2^2\times\ZZ^2	& 0			& \ZZ_2^4			& \ZZ_2^3				& \ZZ_2^6
	\\ \hline
	\mathrm{CI} 	& \ZZ	& \ZZ^4				& 0					& 0					& \ZZ^6		& \ZZ_2^2			& 0					& 0	
	\end{array}
	\end{equation*}
	\caption{Bulk topological invariants and classifying spaces for all AZ symmetry classes with all hexagonal crystal system point group symmetries of crystalline matter, \(C_6,C_{3h},C_{6h},D_6,C_{6v},D_{3h},D_{6h}\). }
\end{table*}

\begin{table*}[t]
	\begin{equation*}
	\begin{array}{l||c||c|c|c|c|c||}
	\multicolumn{2}{l||}{} &
	\parbox{\fgw}{\includegraphics[width=\fgw]{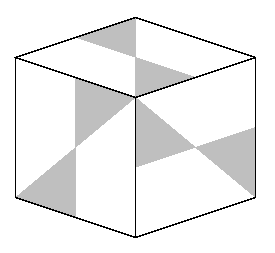}} & 
	\parbox{\fgw}{\includegraphics[width=\fgw]{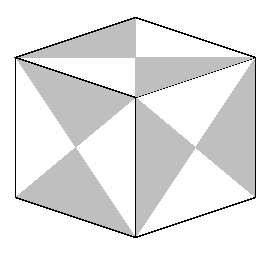}} & 
	\parbox{\fgw}{\includegraphics[width=\fgw]{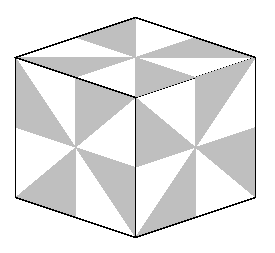}} & 
	\parbox{\fgw}{\includegraphics[width=\fgw]{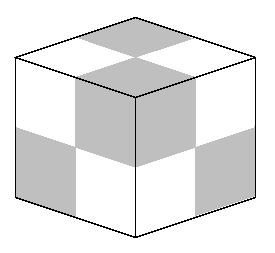}} & 
	\parbox{\fgw}{\includegraphics[width=\fgw]{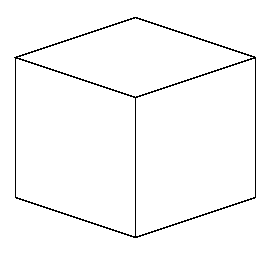}}
	\\ \hline\hline
	\mathrm{Sch\ddot{o}n.}	& C_1 & T & T_h & O & T_d & O_h
	\\ \hline
	\mathrm{HM}	& 1 & 23 & \mathrm{m}\bar{3} & 432 & \bar{4}3\mathrm{m} & \mathrm{m}\bar{3}\mathrm{m}
	\\ \hline\hline
	& \sC_{q+1}	& \sC_{q+1}^4 & \sC_{q}^4 & \sC_{q+1}^5 & \sC_{q}^2\times\sC_{q+1} & \sC_{q}^5
	\\ \hline\hline
	\mathrm{A} 		& 0		& 0					& \ZZ^4				& 0			& \ZZ^2				& \ZZ^5
	\\ \hline
	\mathrm{AIII} 	& \ZZ	& \ZZ^4				& 0					& \ZZ^5		& \ZZ				& 0
	\\ \hline\hline
	& \sR_{q-3}	& \sR_{q-3}^2\times\sC_{q+1} & \sR_{q-4}^2\times\sC_{q} & \sR_{q-3}^5 & \sR_{q-4}^2\times\sR_{q-5} & \sR_{q-4}^5
	\\ \hline\hline
	\mathrm{AI} 	& 0		& 0					& \ZZ^3				& 0			& \ZZ^2				& \ZZ^5
	\\ \hline
	\mathrm{BDI} 	& 0		& \ZZ				& 0					& 0			& \ZZ				& 0
	\\ \hline
	\mathrm{D} 		& 0		& 0					& \ZZ				& 0			& 0					& 0
	\\ \hline
	\mathrm{DIII} 	& \ZZ	& \ZZ^3				& 0					& \ZZ^5		& 0					& 0
	\\ \hline
	\mathrm{AII} 	& \ZZ_2	& \ZZ_2^2			& \ZZ^3				& \ZZ_2^5	& \ZZ^2				& \ZZ^5 
	\\ \hline
	\mathrm{CII} 	& \ZZ_2	& \ZZ_2^2\times\ZZ	& \ZZ_2^2			& \ZZ_2^5	& \ZZ_2^2\times\ZZ	& \ZZ_2^5		
	\\ \hline
	\mathrm{C} 		& 0		& 0					& \ZZ_2^2\times\ZZ	& 0			& \ZZ_2^3			& \ZZ_2^5
	\\ \hline
	\mathrm{CI} 	& \ZZ	& \ZZ^3				& 0					& \ZZ^5		& \ZZ_2				& 0	
	\end{array}
	\end{equation*}
	\caption{Bulk topological invariants and classifying spaces for all AZ symmetry classes with all cubic crystal system point group symmetries of crystalline matter, \(T,T_h,O,T_d,O_h\). }\label{tab:m1}\label{tab:exTd}
\end{table*}

In this paper, we study the structure of the $\ZZ_2$-graded structure algebra extension \({B^0\hookrightarrow B}\) in presence of additional symmetries of the point group, \({g\in G}\), which act on the momenta by
\begin{equation}\label{Og}
U_g^{\phantom{|}}\mathcal{H}(\mathbf{k})U_g^{-1}=\mathcal{H}(O_g\mathbf{k}),
\end{equation}
where \({O_g\in\mathrm{O}(d)}\) is an orthogonal transformation. This yields an action on the Clifford algebras such that \(G\) acts only on the spatial generators
\begin{equation}
U_g^{\phantom{|}}\bm{\gamma}U_g^{-1}=O_g^{\mathrm{T}}\bm{\gamma}.
\end{equation}
However, due to the spin of the electrons, the unitary actions, \(U_g\), form a projective representation, \(\Pi(G)\), satisfying
\begin{equation}\label{spinor}
U_{g}U_{g'}=\pm U_{g\cdot g'}
\end{equation}
(see Sec.~\ref{sec:pgs}). These algebras of spatial and extended non-spatial symmetry actions on the Hilbert space are given by \({B=\langle\Cl_{p+1,q},\Pi(G)\rangle}\), which denotes the algebra generated by the elements of the extended non-spatial symmetry actions, $\Cl_{p+1,q}$, as well as by the projective spatial symmetries, $\Pi(G)$.

We focus on the 32 possible point group symmetries of 3D crystals~\cite{graph}. These are determined by an abstract group structure, \(G\), and an action on 3D space, \({G\subset\mathrm{O}(3)}\).
We identify the algebra extension structure of each point group symmetry within each AZ symmetry class and construct the appropriate classifying space and its bulk topological invariants which are manifested by model Hamiltonians.

In the absence of crystalline symmetry, such bulk topological invariants always manifest in edge modes due to the bulk-boundary correspondence~\cite{franz2013topological,Kane2005Topological,Kane2005Quantum,FuKaneMele2007Topological,Moore2007Topological,Hsieh2009Observation,Roy2009Topological,Fu2007Topological,Konig2007Quantum,xia2009observation,Chen2009Experimental,Khalaf2018Symmetry}. However, in previously treated crystalline symmetries, it was noted that such crystalline bulk invariants may also either host no edge modes~\cite{Hughes2011Inversion,Shiozaki2014Topology}, or in fact host higher-order~\cite{parameswaran2017topological,Benalcazar2017Quantized,Benalcazar2017Electric,Song2017d,Langbehn2017Reflection,Schindler2018Higher,schindler2018bismuth,xu2017topological,Shapourian2018Topological,lin2017topological,Ezawa2018Higher,Khalaf2018Higher,Geier2018Second,trifunovic2018higher,fang2017rotation} hinge or corner protected topological modes. Possible treatment of the higher-order topological insulators and superconductors which correspond to the bulk invariants presented in this paper are discussed in Sec.~\ref{sec:discussion}.

\subsection{Detailed Derivation}\label{sec:analysis}

In this section we provide the mathematical background required for the complete understanding of our technique. The educated reader may skip to the examples in Sec.~\ref{sec:exam}.

\subsubsection{$\mathbb{Z}_2$-Graded Algebras and their Tensor Products}\label{sec:graded}

An algebra $A$ (over $\mathbb{R}$) is called $\mathbb{Z}_2$-graded~\cite{Vela2002Central} if ${A=A^0 \oplus A^1}$ where ${a \cdot a' \in A^{i+j \pmod{2}}}$ for every ${a \in A^i}$ and ${a'\in A^j}$. In this case, $A^0$ is called ``the even part" of $A$, and $A^1$ ``the odd part". The elements of $A^0 \cup A^1$ are called ``homogeneous", the elements of $A^0$ ``even" and the elements of $A^1$ ``odd". The even part forms a subalgebra, \({A^0\hookrightarrow A}\), while the odd part is not an algebra as it is not closed under multiplication.

For example, the Clifford algebra~\cite{kitaev2009periodic,cliff} ${A=\Cl_{p,q}}$, which is the algebra generated over the reals by generators $x_1,\dots,x_p$ and $\gamma_1,\dots,\gamma_q$ which are anticommuting in pairs and satisfy ${x_j^2=-1}$ for every ${j \in \{1,\dots,p\}}$ and ${\gamma_j^2=1}$ for every ${j \in \{1,\dots,q\}}$ is $\mathbb{Z}_2$-graded: with $A^0$ being the usual even part of the Clifford algebra, and $A^1$ being the odd part. The same algebra can admit more than one grading: for example, the algebra of $2\times 2$ real matrices, $M_2(\mathbb{R})$, is generated over $\mathbb{R}$ by $\gamma_1,\gamma_2$ subject to the relations ${\gamma_1^2=\gamma_2^2=1}$ and ${\gamma_1 \gamma_2=-\gamma_2 \gamma_1}$. We can define the $\mathbb{Z}_2$-grading by making both $\gamma_1$ and $\gamma_2$ odd, in which case $\gamma_1 \gamma_2$ will be even, and the even part would be ${\mathbb{R}\langle \gamma_1 \gamma_2\rangle \cong \mathbb{C}}$; this corresponds to the identification \({M_2(\RR)\cong\Cl_{0,2}}\). However, we could also define the grading by making $\gamma_1$ odd and $\gamma_2$ even, in which case the even part would be ${\mathbb{R}\langle \gamma_2 \rangle\cong \mathbb{R}\oplus \mathbb{R}}$; this corresponds to the identification \({M_2(\RR)\cong\Cl_{1,1}}\).

Given two $\mathbb{Z}_2$-graded algebras $A$ and $A'$ generated by homogeneous elements, the graded tensor product $A \hat{\otimes} A'$ is defined to be the $\mathbb{R}$-algebra generated by the generators of $A$ and $A'$ put together such that they satisfy their former relations, plus the following: every even generator of $A$ commutes with all the generators of $A'$, and vice versa, and for odd generators ${a\in A^1}$ and ${a'\in A'^1}$ we have ${aa'=-a'a}$.

Specifically, the \(\ZZ_2\)-graded tensor product~\cite{Vela2002Central} of two Clifford algebras is simply given by
\begin{equation}\label{cliffprod}
\Cl_{p_1,q_1}\hat{\otimes}\Cl_{p_2,q_2}=\Cl_{p_1+p_2,q_1+q_2}.
\end{equation}

Given an algebra $A$, the algebra of $n\times n$ matrices with entries in $A$ is denoted by $M_n(A)$. Recall that by the renowned Artin-Wedderburn theorem, every semi-simple algebra decomposes uniquely as a direct sum of matrix algebras over division algebras. Two semi-simple algebras  are said to be ``Morita equivalent" if they decompose as direct sums of matrix algebras (of possibly different dimensions) over the same division algebras. For example, $\mathbb{R}\oplus \mathbb{C}$ is Morita equivalent to $M_2(\mathbb{R}) \oplus M_3(\CC)$. When the algebras are also $\mathbb{Z}_2$-graded, we say the algebras are equivalent if they are equivalent and also their even parts are Morita equivalent.

\subsubsection{($\ZZ_2$-graded) Algebra Extensions}\label{sec:ext}

By the Bott periodicity~\cite{atiyah1964clifford}, all real Clifford algebras~\cite{cliff} are Morita equivalent to eight prescribed algebras; see Table~\ref{tab:per}.
Their parametrization is a matter of choice, and we follow Kennedy and Zirnbauer~[\onlinecite{kennedy2016bott}] to consider the algebra \({B=\langle\Cl_{q+6+1,d},\Pi(G)\rangle}\) for a given $q \in \{0,1,\dots,7\}$.
The presentation of \(\Cl_{q+6+1,d}\) we consider is the following: write \({\gamma_1,\gamma_2,\dots,\gamma_d}\) for the Dirac gamma matrices, \({x_1,\dots,x_{q+6}}\) for the non-spatial symmetries, and $x_0$ for the spectrally flattened mass matrix. These generators satisfy \({\gamma_i^2=1,~x_i^2=-1}\). We write $y_1,\dots,y_{q+7+d}$ for the generators $x_1,\dots,x_{q+6};x_0;\gamma_1,\dots,\gamma_d$, and we refer to the generators this way in statements which hold for all the generators.

Let us look at a Hilbert space, \(E\), which is a \(B\)-module and hence also a \(B^0\)-module. The choices of Hamiltonians are equivalent to defining a \(B\)-module structure on \(E\) which is compatible with \(B^0\), i.e., all homomorphisms in the category of \(B^0\)-algebras between \(B\) and the endomorphisms of \(E\) as a \(B^0\)-module~\cite{Rowen2008}. This is the definition of the ``\(\Hom\)" functor, \(\Hom_{B^0-\mathbf{alg}}(B,\End_{B^0}{E})\) and we thus associate it with the algebra extension
\begin{equation}\label{MapstoDef}
(B^0\hookrightarrow B)\mapsto \lim_{\dim E\to\infty}\Hom_{B^0-\mathbf{alg}}(B,\End_{B^0}{E}),
\end{equation}
where we take the stable limit~\cite{kitaev2009periodic,kennedy2016bott}. The known results for the Clifford algebra extension problems are thus formulated as
\begin{equation}\label{MapstoR}
\Cl_{p+1,q}\mapsto \sR_{p-q+2},
\end{equation}
where \({\sR_q=\sR_{q+8}}\). Here, we omit the explicit extension \({(\Cl_{p,q}\hookrightarrow\Cl_{p+1,q})}\) since the Clifford algebras always satisfy \({\Cl_{p+1,q}^0=\Cl_{p,q}}\). This classification scheme is reviewed in Appendix~\ref{app:class-space}; see Fig.~\ref{fig:diag1}.

The eightfold Bott periodicity structure~\cite{atiyah1964clifford} in Table~\ref{tab:per} follows from the relations~\cite{cliff}
\begin{equation}
\Cl_{p+1,q+1}\cong M_2(\Cl_{p,q}),\myspace\Cl_{p+8,q}\cong M_{16}(\Cl_{p,q}).
\end{equation}
And indeed, the algebraic structure of Eq.~(\ref{MapstoDef}) is invariant under the Morita equivalence of $B$ with $M_n(B)$.

The complex Clifford algebras~\cite{cliff}, \(\Cl_{q}\), are given by the complexification, \({\CC\otimes_\RR\Cl_{p,q}=\Cl_{p+q}}\), and satisfy
\begin{equation}\label{MapstoC}
\Cl_{q+1}\mapsto \sC_{q},
\end{equation}
where \({\sC_q=\sC_{q+2}}\). Useful identities include,
\begin{equation}\label{RCH}
\HH\otimes_\RR\HH\cong M_4(\RR),\myspace \CC\otimes_\RR\HH\cong M_2(\CC),\myspace \CC\otimes_\RR\CC\cong\CC^\ptwo.
\end{equation}
A particular consequence is that \({\CC\otimes_\RR\Cl_q\cong\Cl_q^\ptwo}\), which enables one to directly read off the classifying spaces of the complex AZ classes (A and AIII) from the classifying spaces of the real AZ classes.

\subsubsection{Point Group Symmetries}\label{sec:pgs}

\def \fgya {41pt}
\begin{figure}[h]
\centering
\begin{tabular}{@{}l@{}c@{}c@{}c@{}c@{}}
& \includegraphics[height=\fgya]{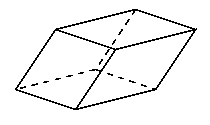} & \includegraphics[height=\fgya]{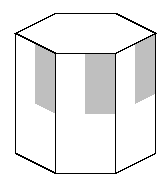} & \includegraphics[height=\fgya]{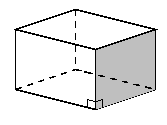} & \includegraphics[height=\fgya]{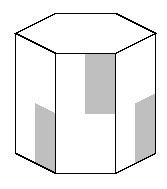}
\\
& Inversion & Rotation & Reflection & Rotoreflection
\\
$G$ & $I^2\!=\!1$ & $(c_n)^n\!=\!1$ & $\sigma^2\!=\!1$ & $(s_{2n})^{2n}\!=\!1$
\\
$\Pi(G)\!\!\!\!\!\!\!\!\!$ & $\hat{I}^2\!=\!1$ & $(\hat{c}_n)^n\!=\!-1$ & $\hat{\sigma}^2\!=\!-1$ & $(\hat{s}_{2n})^{2n}\!=\!-(-1)^{n}$
\end{tabular}
\caption{Solids with either inversion, rotation, reflection, or rotoreflection symmetry. The generators of the symmetry are also specified.}\label{fig:PiG}
\end{figure}

The point groups are generated by: inversions \(I\), rotations \(c_n\) by \(\frac{2\pi}{n}\), reflections \({\sigma=Ic_2}\), and rotoreflections \({s_{2n}=c_{2n}\sigma_h}\), where \(\sigma_h,\sigma_v\) are horizontal and vertical reflections.

Importantly, one must take into account the fermionic nature of the electrons and hence the projective spinor representation $\Pi(G)$; see Eq.~(\ref{spinor}). This spinor representation is constructed from subgroups of the Pin group, which is a double cover of the orthogonal group [just as the Spin group, ${\mathrm{Spin}(3)\cong\mathrm{SU}(2)}$, is a double cover of the special orthogonal group]:
\begin{equation}
\begin{array}{ccccc}
\{\pm1\} & \hookrightarrow & \mathrm{SU}(2) & \twoheadrightarrow & \mathrm{SO}(3)
\\
\veq & & \hookdownarrow & & \hookdownarrow
\\
\{\pm1\} & \hookrightarrow & \mathrm{Pin}_-(3) & \twoheadrightarrow & \mathrm{O}(3)
\\
 & & \twoheaddownarrow & & \twoheaddownarrow
\\
&  & \{\pm1\} & = & \{\pm1\}
\end{array}
\end{equation}
Within the Pin (and Spin) group, one obtains the fermionic property that a $4\pi$ rotation equals 1. The subgroups are known as double point groups, \({\hat{G}\subset\mathrm{Pin}_-(3)}\), such that \({\hat{G}/\ZZ_2=G}\).
However, when taking into account the projective nature of the Hilbert space and constructing the spinor representation, this amounts to setting a \(2\pi\) rotation to be \({(\hat{c}_n)^n=-1}\), and a double inversion to be \({\hat{I}^2=1}\). This is referred to as the spinor representation, \(\Pi(G)\). 
To clarify things~\cite{Ando2015Topological}, we note that:\\
(i)\phantom{ii} In \(G\) a \(2\pi\) rotation is \(1\).\\
(ii)\phantom{i} In \(\hat{G}\) a \(2\pi\) rotation squares to \(1\).\\
(iii) In \(\Pi(G)\) a \(2\pi\) rotation is \(-1\).

For brevity, we shall use \(\hat{c},\hat{s},\hat{\sigma}\) for the generators of both \(\hat{G}\) and \(\Pi(G)\). Moreover, one finds that a reflection satisfies \({\hat{\sigma}^2=I^2\hat{c}_2^2=-1}\), and that since the product of orthogonal reflections \(\hat\sigma_h\hat\sigma_v\) is a \(\pi\) rotation, they satisfy \({(\hat\sigma_h\hat\sigma_v)^2=-1}\). The rotoreflections, however, depend on the parity, as \({(\hat s_{2n})^{2n}=\hat c_{2n}^{2n}(\hat\sigma_h^2)^n=-(-1)^n}\) (see Fig.~\ref{fig:PiG}).

The point group symmetries each correspond to one of the abstract groups $\ZZ_n$, $\Dih_n$, $A_4$, $S_4$, or their product with $\ZZ_2$, as presented in Table~\ref{tab:pgs}.
Useful identities include,
\begin{equation}\label{prodgroup}
\ZZ_{6}=\ZZ_{3}\times\ZZ_2,\myspace\Dih_{2}=\ZZ_2\times\ZZ_2,\myspace\Dih_{6}=\Dih_{3}\times\ZZ_2.
\end{equation}

\begin{table}[t]
\begin{tabular}{l|c||l|c}
\multicolumn{2}{c||}{Abstract groups} & \multicolumn{2}{c}{Point group symmetries}\\
\hline
Name & Symbol & Name & Sch\"{o}n.\\
\hline\hline
Cyclic & \(\ZZ_n\) & Rotational & $C_n$\\
& $\ZZ_{2n}$ & Rotoreflection & $S_{2n}$\\
\hline
Dihedral & \(\Dih_n\) & Pyramidal & $C_{nv}$\\
& \(\Dih_n\) & Dihedral & $D_{n}$\\
& $\Dih_{2n}$ & Antiprismatic & $D_{nd}$ \\
\hline
Alternating & \(A_4\) & Chiral tetrahedral & $T$\\
\hline
Symmetric & \(S_4\) & Full tetrahedral & $T_d$\\
& \(S_4\) & Chiral octahedral & $O$\\
\hline
& $\ZZ_n\times\ZZ_2$ & Dipyramidal & $C_{nh}$\\
\cline{2-4}
& $\Dih_n\times\ZZ_2$ & Prismatic & $D_{nh}$\\
\cline{2-4}
& $A_4\times\ZZ_2$ & Pyritohedral & $T_h$\\
\cline{2-4}
& $S_4\times\ZZ_2$ & Full octahedral & $O_h$\\
\end{tabular}
\caption{The abstract groups corresponding to each of the point group symmetries.}\label{tab:pgs}
\end{table}

%
%
%
%
%
%
%

\subsubsection{$\mathbb{Z}_2$-Graded Groups and their Representations}\label{sec:ring}

A group $G$ is called $\mathbb{Z}_2$-graded~\cite{vershik2008new} if there exists a group homomorphism \({G\to\ZZ_2}\), in this case, the kernel of the homomorphism is denoted $G^0$ and dubbed ``the even part", while ${G^1=G\setminus G^0}$ is dubbed ``the odd part". Similar to Sec.~\ref{sec:graded}, one has ${G=G^0 \cup G^1}$, where ${g \cdot g' \in G^{i+j \pmod{2}}}$ for every ${g \in G^i}$ and ${g'\in G^j}$. The even part forms a normal subgroup, \({G^0\triangleleft G}\), while the odd part is not a group as it is not closed under multiplication.

The simplest case is when $G\xrightarrow{0}\ZZ_2$, such that all element are even and ${G^0=G}$.

Otherwise, $G\twoheadrightarrow\ZZ_2$, and by the ``first isomorphism theorem", ${G^0\triangleleft G}$ is a normal subgroup and
\begin{equation}
G/G^0=\ZZ_2.
\end{equation}

A $\ZZ_2$-graded representation~\cite{vershik2008new} of a $\ZZ_2$-graded group, $G$, is a representation, $\rho$, into $\ZZ_2$-graded matrix algebras that preserve parity, i.e., the even/odd elements of ${G=G^0 \cup G^1}$ are represented by even/odd elements of ${A_\rho=A_\rho^0 \oplus A_\rho^1}$. 

When studying group representations, a key notion is that of the ``group ring"~\cite{Passman1977}. It is 
the algebra generated by a group $G$ over $\RR$ (or over any other ring) and denoted $\RR[G]$. It is defined to be the $\RR$-algebra, $\bigoplus_{g\in G} \RR g$, whose basis as a real vector space consists of the elements of $G$, and its multiplication table is given by $(r_1 g_1)(r_2 g_2)=(r_1 r_2)(g_1 g_2)$. 
Moreover, by construction,
\begin{equation}
G^0\triangleleft G \myspace\Rightarrow\myspace \RR[G^0]\hookrightarrow\RR[G],
\end{equation}
and a $\ZZ_2$ grading of $G$ induces a $\ZZ_2$ grading of $\RR[G]$ such that $\RR[G^0]$ forms the even part of $\RR[G]$, i.e., ${\RR[G]^0=\RR[G^0]}$ and ${\RR[G]^1=\bigoplus_{g\in G^1} \RR g}$.

A highly useful property, which follows the Maschke and Artin-Wedderburn theorems, is that the group ring $\RR[G]$ is isomorphic to a direct sum of the matrix algebras corresponding to each irreducible representation of $G$.

We thus list the group rings of all point group symmetries:

Write \({n=2c(n)+r(n)}\), where
\begin{equation}
c(n)=\left\lfloor\frac{n-1}{2}\right\rfloor,\myspace
r(n)=\begin{cases}
1 & n~\mathrm{odd},\\
2 & n~\mathrm{even}.
\end{cases}
\end{equation}
The group rings, associated with the abstract groups corresponding to the point group symmetries in Table~\ref{tab:pgs}, are each equivalent to one of the following cases,
\begin{equation}
\begin{aligned}
\RR[\ZZ_n]&=\RR^{\oplus r(n)}\oplus\CC^{\oplus c(n)},\\
\RR[\Dih_n]&=(\RR^\ptwo)^{\oplus r(n)}\oplus M_2(\RR)^{\oplus c(n)},\\
\RR[A_4]&=\RR\oplus\CC\oplus M_3(\RR),\\
\RR[S_4]&=\RR^\ptwo\oplus M_2(\RR)\oplus M_3(\RR)^\ptwo,\\
\RR[G\times\ZZ_2]&=\RR[G]^\ptwo.
\end{aligned}
\end{equation}

\subsubsection{From Graded Tensor Products to Topological Invariants}\label{rgtp}

By successively analysing all possible point group symmetries, we show that within the extended algebra of symmetries, \({B=\langle\Cl_{p+1,q},\Pi(G)\rangle}\), one can always make a change of variables and \textit{redefine the action} of $G$ as a $\ZZ_2$-graded abstract group, ${G=G^0\cup G^1}$ (see Appendix~\ref{app:pgf}). We find elements of $B$ which satisfy ${\tilde{U}_{g}\tilde{U}_{g'}=\tilde{U}_{g\cdot g'}}$ such that the even part, $G^0$, acts trivially on \(\Cl_{p+1,q}\); the odd elements, ${g\in G^1}$, act by ${\tilde{U}_g^{\phantom{1}}a\tilde{U}_g^{-1}=-a}$ on the odd elements of the Clifford algebra, ${a\in\Cl_{p+1,q}^1}$.

These algebraic relations, by definition, bring the extended algebra of symmetries to a $\ZZ_2$-graded tensor product structure, \({B=\Cl_{p+1,q}\hat\otimes\RR[G]}\), (see Sec.~\ref{sec:graded}) where the grading of $\RR[G]$ is induced by $G$. This presentation is much simpler than the original projective action, $\Pi(G)$, (see Sec.~\ref{sec:paradigm}) since the structure of $B$ is now determined by identifying the \(\ZZ_2\)-grading of $G$.

If \(G\) acts trivially on \(\Cl_{p+1,q}\) then we have the simplest cases  of \({B=\RR[G]\otimes\Cl_{p+1,q}}\). 
Otherwise, we must have a nontrivial grading
\begin{equation}
G/G^0=\ZZ_2
\end{equation}
(see Sec.~\ref{sec:ring}).
In order to determine the $\ZZ_2$-graded structure of the group ring corresponding to a specific point group symmetry, we use either ${G=G^0\times\ZZ_2}$ or one of the following possible gradings:
\begin{equation}
\begin{gathered}
\ZZ_{2n}/\ZZ_{n}=\ZZ_2,\myspace\Dih_{2n}/\Dih_{n}=\ZZ_2,\\
\Dih_n/\ZZ_n=\ZZ_2,\myspace S_4/A_4=\ZZ_2.\\
\end{gathered}
\end{equation}
The grading is determined by identifying the correct even normal subgroup, \({G^0\triangleleft G}\), acting trivially on \(\Cl_{p+1,q}\).

This enables us to decompose the group ring $\RR[G]$ according to irreducible $\ZZ_2$-graded representations of $G$,
\begin{equation}
\RR[G]=\bigoplus\nolimits_{\rho} A_{\rho},\myspace
\RR[G^0]=\bigoplus\nolimits_{\rho} A^0_{\rho},
\end{equation}
where $A^0_{\rho}$ is the even part of $A_{\rho}$, see Sec.~\ref{sec:ring}. Note, that for each $\ZZ_2$-graded representation, $\rho$, either $A_\rho$ is an \textit{ungraded} representation of $G$, or $A_\rho^0$ is an \textit{ungraded} representation of $G^0$. Hence, these representations may be used to label~\cite{cotton2003chemical} $\rho$ (this is demonstrated in Sec.~\ref{sec:Td}). 

Since any $\ZZ_2$-graded real algebra is Morita equivalent to a direct sum of Clifford algebras, we may always write
\begin{equation}
\RR[G]=\bigoplus\nolimits_{\rho^\RR}[M_n(\Cl_{p'',q''})]_{\rho^\RR}\oplus\bigoplus\nolimits_{\rho^\CC}[M_n(\Cl_{q''})]_{\rho^\CC},
\end{equation}
where ${\{\rho^{\RR}\}\cup\{\rho^{\CC}\}=\{\rho\}}$ denote the irreducible $\ZZ_2$-graded representations for which $A_{\rho}$ is either real or complex as a $\ZZ_2$-graded algebra. 
This decomposition enables us to identify the algebraic structure of the extended algebra of symmetries, and we thus find a Clifford algebra structure corresponding to each irreducible $\ZZ_2$-graded representation
\begin{align}
B&=\Cl_{p+1,q}\hat\otimes\RR[G]\\
&=\bigoplus\nolimits_{\rho^\RR}[M_n(\Cl_{p'+1,q'})]_{\rho^\RR}\oplus\bigoplus\nolimits_{\rho^\CC}[M_n(\Cl_{q'+1})]_{\rho^\CC}.\nonumber
\end{align}
Here, using the algebraic properties of Clifford algebras presented in Secs.~\ref{sec:graded} and \ref{sec:ext}, we find \({p'=p+p''}\) and \({q'=q+q''}\) for the real representations, and \({q'=p+q+q''}\) for the complex representations. This is demonstrated in Sec.~\ref{sec:exam}. 

In accordance with Sec.~\ref{sec:ext}, the Clifford algebras corresponding to each $\ZZ_2$-graded representation $\rho$ maps to  either a real or a complex classifying space,
\begin{equation}
B\mapsto\prod\nolimits_{\rho^\RR}[\sR_{p'-q'+2}]_{\rho^\RR}\times\prod\nolimits_{\rho^\CC}[\sC_{q'}]_{\rho^\CC}.
\end{equation}
The topological indices classifying materials with point group symmetry $G$, presented in Tables~\ref{tab:p1}-\ref{tab:m1}, are the topological invariants of these classifying spaces.

Let us denote a Hamiltonian~\cite{Schnyder2008Classification,ryu2010topological} of AZ class $q$ in $d$ spatial dimensions by $\mathcal{H}_{q,d}^{\RR}$ for the eight real AZ classes, and by $\mathcal{H}_{q,d}^{\CC}$ for the two complex AZ classes.

The Hamiltonian, $\mathcal{H}^{\RR,G}_{q,d}$, of a crystalline material with point group symmetry $G$ in real AZ class $q$, may be block-decomposed into irreducible $\ZZ_2$-graded representations $\{\rho\}$ of $G$:
\begin{equation}\label{blocks}
\mathcal{H}^{\RR,G}_{q,d}=\bigoplus\nolimits_{\rho^\RR}[\mathcal{H}^{\RR}_{p'-q'+2+d,d}]_{\rho^\RR}\oplus\bigoplus\nolimits_{\rho^\CC}[\mathcal{H}^{\CC}_{q'+d,d}]_{\rho^\CC}.
\end{equation}
Each block is equivalent to an AZ Hamiltonian which is determined by the Clifford algebras 
corresponding to the irreducible $\ZZ_2$-graded representation, $\rho$. Note, that a simpler analogous construction applies for the Hamiltonians, $\mathcal{H}^{\CC,G}_{q,d}$, of complex AZ classes.

The standard topological invariants~\cite{Schnyder2008Classification,ryu2010topological,Shiozaki2014Topology,Chiu2016Classification} characterize the block Hamiltonians of each irreducible $\ZZ_2$-graded representation. These invariants are presented in Table~\ref{tab:inv}.

\begin{table}[t]
\begin{tabular}{l|c||c||c|c}
& $s$ & & $d$ even & $d$ odd
\\ \hline\hline
complex
& 0 & $\ZZ$ & Chern number & winding number
\\ \hline\hline
real
& 0,4 & $\ZZ$ & Chern number & winding number
\\ \cline{2-5}
& 1,2 & $\ZZ_2$ & Fu-Kane invariant & Chern-Simons integral
\end{tabular}
\caption{Topological invariants~\cite{Schnyder2008Classification,ryu2010topological,Shiozaki2014Topology,Chiu2016Classification} characterizing the Hamiltonians, $[\mathcal{H}^{\RR/\CC}_{s+d,d}]_{\rho}$, corresponding to irreducible $\ZZ_2$-graded representations; see Eq.~(\ref{blocks}).\label{tab:inv}}
\end{table}

\begin{table*}[t]
	\begin{align*}
	&\begin{array}{|l|l||c|c|c||c|c|c||c|c|c}
	\hline
	\multicolumn{2}{|l||}{\mathrm{Sch\ddot{o}nflies}} & C_1 & C_{(2n)} & C_3 & C_{1h} & C_{(2n)h} & C_{3h} & C_{1v} & C_{(2n)v} & C_{3v}
	\\ \hline
	\multicolumn{2}{|l||}{\mathrm{HM}} & 1 & (2n) & 3 & \mathrm{m} & (2n)/\mathrm{m} & 3/\mathrm{m} & \mathrm{m} & (2n)\mathrm{mm} & 3\mathrm{m}
	\\ \hline\hline
	\mathrm{3D} & \mathrm{complex}
	& \sC_{q+1}	& \sC_{q+1}^{2n} & \sC_{q+1}^3 & \sC_{q}	& \sC_{q}^{2n} & \sC_{q}^3 & \sC_{q}	& \sC_{q}^2\!\times\!\sC_{q+1}^{n-1} & \sC_{q}\!\times\!\sC_{q+1}
	\\ \cline{2-11}
	& \mathrm{real}
	& \sR_{q-3}	& \sR_{q-3}^2\!\times\!\sC_{q+1}^{n-1} & \sR_{q-3}\!\times\!\sC_{q+1} & \sR_{q-4}	& \sR_{q-4}^2\!\times\!\sC_{q}^{n-1} & \sR_{q-4}\!\times\!\sC_{q} & \sR_{q-4}	& \sR_{q-4}^2\!\times\!\sR_{q-5}^{n-1} & \sR_{q-4}\!\times\!\sR_{q-5}
	\\ \hline\hline
	\mathrm{2D} & \mathrm{complex}
	& \sC_{q}	& \sC_{q}^{2n} & \sC_{q}^3 & \sC_q^2 & \sC_q^{4n} & \sC_q^6 & \sC_{q+1}	& \sC_{q+1}^2\!\times\!\sC_{q}^{n-1} & \sC_{q+1}\!\times\!\sC_{q}
	\\ \cline{2-11}
	& \mathrm{real}
	& \sR_{q-2}	& \sR_{q-2}^2\!\times\!\sC_{q}^{n-1} & \sR_{q-2}\!\times\!\sC_{q} & \sC_q & \sC_q^{2n} & \sC_q^3 & \sR_{q-3}	& \sR_{q-3}^2\!\times\!\sR_{q-4}^{n-1} & \sR_{q-3}\!\times\!\sR_{q-4}
	\\\hline
	\end{array}\\\\
	&\begin{array}{l|l||c|c|c||c|c|c||c|c|c}
	\hline
	\multicolumn{2}{l||}{\mathrm{Sch\ddot{o}nflies}} & D_1 & D_{(2n)} & D_3 & D_{1h} & D_{(2n)h} & D_{3h} & D_{1d} & D_{2d} & D_{3d}
	\\ \hline
	\multicolumn{2}{l||}{\mathrm{HM}} & 2 & (2n)22 & 32 & \mathrm{mm}2 & (2n)/\mathrm{mmm} & \bar{6}\mathrm{m}2 & 2/\mathrm{m} & \bar{4}2\mathrm{m} & \bar{3}\mathrm{m}
	\\ \hline\hline
	\mathrm{3D} & \mathrm{complex}
	& \sC_{q+1}^2 & \sC_{q+1}^{n+3} & \sC_{q+1}^{3} & \sC_{q}^2 & \sC_{q}^{n+3} & \sC_{q}^{3}	& \sC_{q}^{2} & \sC_{q}^{2}\!\times\!\sC_{q+1} & \sC_{q}^{3}
	\\ \cline{2-11}
	& \mathrm{real}
	& \sR_{q-3}^2 & \sR_{q-3}^{n+3} & \sR_{q-3}^{3} & \sR_{q-4}^2 & \sR_{q-4}^{n+3} & \sR_{q-4}^{3}	& \sR_{q-4}^{2} & \sR_{q-4}^{2}\!\times\!\sR_{q-3} & \sR_{q-4}^{3}
	\\ \hline\hline
	\mathrm{2D} & \mathrm{complex}
	& \sC_{q+1}	& \sC_{q+1}^2\!\times\!\sC_{q}^{n-1} & \sC_{q+1}\!\times\!\sC_{q} & \sC_{q} & \sC_{q}^{2n} & \sC_{q}^3 & \sC_{q} & \sC_{q+1}^2\!\times\!\sC_{q} & \sC_q^3
	\\ \cline{2-11}
	& \mathrm{real}
	& \sR_{q-3}	& \sR_{q-3}^2\!\times\!\sR_{q-4}^{n-1} & \sR_{q-3}\!\times\!\sR_{q-4} & \sR_{q-2} & \sR_{q-2}^{2n} & \sR_{q-2}^3 & \sR_{q-4} & \sR_{q-3}^2\!\times\!\sR_{q-4} & \sR_{q-4}\!\times\!\sC_q
	\\\hline
	\end{array}\\\\
	&\begin{array}{l|l||c|c|c||c|c|c|c|c|}
	\hline
	\multicolumn{2}{l||}{\mathrm{Sch\ddot{o}nflies}} & S_2 & S_4 & S_6 & T & T_h & T_d & O & O_h	
	\\ \hline
	\multicolumn{2}{l||}{\mathrm{HM}} & \bar{1} & \bar{4} & \bar{3} & 23 & \mathrm{m}\bar{3} & \bar{4}3\mathrm{m} & 432 & \mathrm{m}\bar{3}\mathrm{m}
	\\ \hline\hline
	\mathrm{3D} & \mathrm{complex} 
	& \sC_{q} & \sC_{q}^2 & \sC_{q}^3 & \sC_{q+1}^{4} &  \sC_{q}^{4} & \sC_{q}^2\!\times\!\sC_{q+1} & \sC_{q+1}^{5} & \sC_{q}^{5}
	\\ \cline{2-10}
	& \mathrm{real} 
	& \sR_{q-4} & \sR_{q-4}\!\times\!\sR_{q-2} & \sR_{q-4}\!\times\!\sC_{q} & \sR_{q-3}^{2}\!\times\!\sC_{q+1} &  \sR_{q-4}^{2}\!\times\!\sC_{q} & \sR_{q-4}^2\!\times\!\sR_{q-5} & \sR_{q-3}^{5} & \sR_{q-4}^{5}
	\\ \hline\hline
	\mathrm{2D} & \mathrm{complex} 
	& \sC_q^2 & \sC_{q}^{4} & \sC_q^6
	\\ \cline{2-5}
	& \mathrm{real} 
	& \sC_q & \sR_{q-2}^{2}\!\times\!\sC_{q} & \sC_q^3
	\\\cline{1-5}
	\end{array}
	\end{align*}
	\caption{Classifying spaces of all point group symmetries of 3D crystals and all symmorphic layer group symmetries of 2D crystals.}
	\label{tab:class}
\end{table*}

%

Our classification scheme, which was presented in this section, is fully demonstrated within the following concrete examples.

\section{Examples}\label{sec:exam}
We now give three pedagogical examples demonstrating our technique. We then use these examples to demonstrate the construction of a model Hamiltonian manifesting our classification results. The analyses for all other point group symmetries are brought in Appendix~\ref{app:pgf}. 

\subsection{Threefold Rotational Symmetry \(C_3\)}\label{sec:c3}
\begin{figure}[h]
\centering
\includegraphics[height=\fgx]{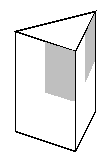}
\caption{A solid with the threefold rotational symmetry point group $C_3$ of trigonal-pyramidal crystals.}
\end{figure}

Let us begin by considering a crystalline insulator (or superconductor) with a threefold rotational symmetry point group $C_3$ which is one of the simplest point group symmetries. As an abstract group, the symmetry group is given by the cyclic group \({G=\ZZ_3}\), and we mark the generator of its projective fermionic representation by \(\hat{c}\) such that
\begin{equation}
\hat c^3=-1.
\end{equation}
The action of this generator on the 3D space is given by a simple threefold rotation
\begin{equation}
\hat c(\begin{smallmatrix}\gamma_1\\\gamma_2\end{smallmatrix})\hat c^{-1}=(\begin{smallmatrix}\cos\frac{2\pi}{3}&\sin\frac{2\pi}{3}\\-\sin\frac{2\pi}{3}&\cos\frac{2\pi}{3}\end{smallmatrix})(\begin{smallmatrix}\gamma_1\\\gamma_2\end{smallmatrix}).
\end{equation}

We wish to find the structure of the algebra \({B=\langle\Cl_{q+7,3},\Pi(\ZZ_3)\rangle}\) (and corresponding classifying space) generated by $\hat{c}$ as well as by the Dirac gamma matrices, $\gamma_1,\gamma_2,\gamma_3$, and the other non spatial Clifford algebra generators \(\{\gamma\}\subset\{y\}\). In order to do so, we notice that
\begin{equation}\label{rotrel}
\begin{aligned}
e^{-\gamma_1\gamma_2\frac{2\pi}{6}}(\begin{smallmatrix}\gamma_1\\\gamma_2\end{smallmatrix})e^{\gamma_1\gamma_2\frac{2\pi}{6}}
&=e^{-\gamma_1\gamma_2\frac{2\pi}{3}}(\begin{smallmatrix}\gamma_1\\\gamma_2\end{smallmatrix}) \\ &=(\begin{smallmatrix}\cos\frac{2\pi}{3}&\sin\frac{2\pi}{3}\\-\sin\frac{2\pi}{3}&\cos\frac{2\pi}{3}\end{smallmatrix})(\begin{smallmatrix}\gamma_1\\\gamma_2\end{smallmatrix}),
\end{aligned}
\end{equation}
where we used the anticommutation relations of the Dirac gamma matrices; see Eq.~(\ref{gammapt}). This prompts the definition of a new abstract generator $c$ such that
\begin{equation}
c = e^{\gamma_1\gamma_2\frac{2\pi}{6}}\hat c.
\end{equation}
This new abstract generator satisfies
\begin{equation}\label{c3rel}
\myspace c y_i c^{-1}=y_i,\myspace c^3=1.
\end{equation}
We see that $c$ is completely decoupled from the generators of the Clifford algebra, $y_i$, and thus get \({B=\RR[\ZZ_3]\otimes\Cl_{q+7,3}}\) and hence
\begin{equation}\label{resc3}
(\RR\oplus\CC)\otimes\Cl_{q+7,3}\mapsto\sR_{q-3}\times\sC_{q+1}.
\end{equation}
Here, we have used the mapping in Sec.~\ref{sec:ext} to determine the classifying space.

The classifying spaces of the complex AZ classes immediately follow by utilizing ${\CC\otimes\Cl_{p,q}=\Cl_{p+q}}$ and ${\CC\otimes\Cl_{q}=\Cl_{q}^\ptwo}$,
\begin{equation}\label{resc3c}
(\RR\oplus\CC)\otimes\Cl_{q+10}\mapsto\sC_{q+1}^3.
\end{equation}

The topological indices classifying materials with $C_3$ threefold rotational symmetry, presented in Table~\ref{tab:exc3}, are the topological invariants of these classifying spaces [Eqs.~(\ref{resc3}) and (\ref{resc3c})].

\subsubsection*{Model Hamiltonians and Topological Invariants}
In order to construct a model Hamiltonian for these crystalline phases, one can use the model Hamiltonians of the tenfold-way classification with non-spatial symmetries~\cite{Schnyder2008Classification,ryu2010topological}. We denote the model Hamiltonian of AZ class $q$ in $d$ spatial dimensions by $\mathcal{H}_{q,d}^{\RR}$ for the eight real AZ classes, and by $\mathcal{H}_{q,d}^{\CC}$ for the two complex AZ classes; see Sec.~\ref{rgtp}.

As a consequence of Eq.~(\ref{c3rel}) one can diagonalize the Hamiltonians for $C_3$ threefold rotational symmetry simultaneously with $c$. We thus write a model Hamiltonian, $\mathcal{H}_{q,3}^{\RR,C_3}$, in a block diagonal form, labelled by the eigenvalues of $c$:
\begin{equation}
\mathcal{H}_{q,3}^{\RR,C_3}=\begin{pmatrix}
\big[\mathcal{H}_{q,3}^{\RR}\big]_{c=1} & 0 \\
0 & \big[\mathcal{H}_{q,3}^{\CC}\big]_{c=e^{\pm\frac{2\pi i}{3}}}
\end{pmatrix}.
\end{equation}
The complex AZ class yield a similar decomposition,
\begin{equation}
\mathcal{H}_{q,3}^{\CC,C_3}=\begin{pmatrix}
\big[\mathcal{H}_{q,3}^{\CC}\big]_{c=1} & 0 & 0 \\
0 & \big[\mathcal{H}_{q,3}^{\CC}\big]_{c=e^{+\frac{2\pi i}{3}}} & 0 \\
0 & 0 & \big[\mathcal{H}_{q,3}^{\CC}\big]_{c=e^{-\frac{2\pi i}{3}}}
\end{pmatrix}.
\end{equation}
The topological invariants of each block are the topological invariants of the corresponding Hamiltonians~\cite{Schnyder2008Classification,ryu2010topological,Shiozaki2014Topology,Chiu2016Classification}; see Table~\ref{tab:inv}.

\subsection{Fourfold Rotoreflection Symmetry \(S_{4}\)}\label{sec:s4}
\begin{figure}[h]
\centering
\includegraphics[height=\fgx]{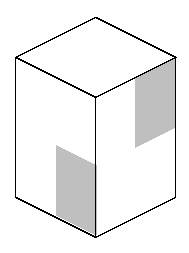}
\caption{A solid with the fourfold rotoreflection symmetry point group \(S_4\) of tetragonal-disphenoidal crystals.\label{fig:s4}}
\end{figure}

Let us next consider a crystalline insulator (or superconductor) with a fourfold rotoreflection symmetry point group $S_4$. As an abstract group, the symmetry group is given by the cyclic group \({G=\ZZ_4}\), and we mark the generator of its projective fermionic representation by \(\hat{s}_4\) such that
\begin{equation}
\hat s^{4}=-1.
\end{equation}
The action of this generator on the 3D space is given by a fourfold rotoreflection
\begin{equation}
\hat s(\begin{smallmatrix}\gamma_1\\\gamma_2\end{smallmatrix})\hat s^{-1}=(\begin{smallmatrix}0&1\\-1&0\end{smallmatrix})(\begin{smallmatrix}\gamma_1\\\gamma_2\end{smallmatrix}),\myspace \hat s\gamma_3 \hat s^{-1}=-\gamma_3.
\end{equation}
We wish to find the structure of the algebra \({B=\langle\Cl_{q+7,3},\Pi(\ZZ_4)\rangle}\) (and corresponding classifying space) generated by $\hat{s}_4$ as well as by the Dirac gamma matrices, $\gamma_1,\gamma_2,\gamma_3$, and the other non spatial Clifford algebra generators \(\{\gamma\}\subset\{y\}\). In an analogous manner to Eq.~(\ref{rotrel}) we define new abstract generators \(s_4\) such that
\begin{equation}
s = e^{\gamma_1\gamma_2\frac{\pi}{4}}\hat s\gamma_3.
\end{equation}
This new abstract generator satisfies
\begin{equation}\label{s4rel}
s y_i s^{-1}= -y_i,\myspace s^{4}=1.
\end{equation}
Here, we used the equality \({(s^{2})^2=(-e^{-\gamma_1\gamma_2\frac{2\pi}{4}}\hat s^2)^2=1}\).

Trying to analyse the algebraic structure, we find a central element \({t=s^2}\) satisfying ${t^2=1}$. Our algebra decomposes accordingly as a direct sum of two algebras, such that in one of them \({s^2=t=1}\) and in the other \({s^2=t=-1}\). In each of these algebras, the appropriate extra generator (${s^2=\pm1}$) is simply added to the generators, $y_i$, of the original Clifford algebra, $\Cl_{q+7,3}$, and we therefore get
\begin{equation}
\Cl_{q+7,3+1}\oplus\Cl_{q+7+1,3}\mapsto\sR_{q-4}\times\sR_{q-2},
\end{equation}
where we have once again used the mapping in Sec.~\ref{sec:ext} to determine the classifying space.

The classifying spaces of the complex AZ classes immediately follow by utilizing $\CC\otimes\Cl_{p,q}=\Cl_{p+q}$:
\begin{equation}\label{ress4c}
\Cl_{q+11}\oplus\Cl_{q+11}\mapsto\sC_{q}^2.
\end{equation}
Moreover, the same results may be achieved by noticing that the relations in Eq.~(\ref{s4rel}) are exactly the relations of a $\ZZ_2$-graded tensor product \({B=\Cl_{q+7,3}\hat{\otimes}\RR[\ZZ_4]}\) where the ``even" part of $\RR[\ZZ_4]$, which commutes with the generators of $\Cl_{q+7,3}$, is generated by elements with the generator, $s$, occurring an even number of  times, i.e., $s^2$ and $1$.
These even elements generate the subalgebra $\RR[\ZZ_2]$ of the cyclic group \({\ZZ_2\triangleleft \ZZ_4}\). By knowing the grading of the groups \({\ZZ_4/\ZZ_2=\ZZ_2}\), we can find the grading of the group rings \({\RR[\ZZ_2]\hookrightarrow\RR[\ZZ_4]}\) as a direct sum of Clifford algebras
\begin{equation}
\begin{matrix}
\RR[\ZZ_2]&=&\RR\oplus\RR\\
\hookdownarrow&&\hookdownarrow&&\\
\RR[\ZZ_4]&=&\RR^\ptwo\oplus\CC&=&\Cl_{0,1}\oplus\Cl_{1,0}.
\end{matrix}
\end{equation}
We can thus find its graded tensor product with any Clifford algebra
\begin{equation}
\begin{aligned}
\Cl_{p,q}\hat{\otimes}\RR[\ZZ_4]&=\Cl_{p,q}\hat{\otimes}(\Cl_{0,1}\oplus\Cl_{1,0})\\&=\Cl_{p,q+1}\oplus\Cl_{p+1,q}.
\end{aligned}
\end{equation}
And specifically for \({B=\Cl_{q+7,3}\hat{\otimes}\RR[\ZZ_4]}\) we find
\begin{equation}\label{ress4}
\Cl_{q+7,3+1}\oplus\Cl_{q+7+1,3}\mapsto\sR_{q-4}\times\sR_{q-2}.
\end{equation}
Such analyses would come in handy in other more complicated cases such as our final example.

The topological indices classifying materials with $S_4$ fourfold rotoreflection symmetry, presented in Table~\ref{tab:exs4}, are the topological invariants of these classifying spaces, Eqs.~(\ref{ress4}),~(\ref{ress4c}).

\subsubsection*{Model Hamiltonians and Topological Invariants}
In order to construct a model Hamiltonian for these crystalline phases, one can use the model Hamiltonians of the tenfold-way classification with non-spatial symmetries~\cite{Schnyder2008Classification,ryu2010topological}, $\mathcal{H}_{q,d}^{\RR/\CC}$, for real/complex AZ class $q$ in $d$ spatial dimensions; see Sec.~\ref{rgtp}.

As a consequence of Eq.~(\ref{s4rel}) one can diagonalize the Hamiltonians for $S_4$ fourfold rotoreflection symmetry simultaneously with $s^2$. We thus write a model Hamiltonian, $\mathcal{H}_{q,3}^{\RR,S_4}$, in a block diagonal form, labelled by the eigenvalues of $s^2$:
\begin{equation}
\mathcal{H}_{q,3}^{\RR,S_4}=\begin{pmatrix}
\big[\mathcal{H}_{q-1,3}^{\RR}\big]_{s^2=+1} & 0 \\
0 & \big[\mathcal{H}_{q+1,3}^{\RR}\big]_{s^2=-1}
\end{pmatrix}.
\end{equation}
The complex AZ class yield a similar decomposition,
\begin{equation}
\mathcal{H}_{q,3}^{\CC,S_4}=\begin{pmatrix}
\big[\mathcal{H}_{q+1,3}^{\CC}\big]_{s^2=+1} & 0 \\
0 & \big[\mathcal{H}_{q+1,3}^{\CC}\big]_{s^2=-1}
\end{pmatrix}.
\end{equation}
The topological invariants of each block are the topological invariants of the corresponding Hamiltonians~\cite{Schnyder2008Classification,ryu2010topological,Shiozaki2014Topology,Chiu2016Classification}; see Table~\ref{tab:inv}.

\subsection{Full Tetrahedral Symmetry \(T_d\)}\label{sec:Td}
\begin{figure}[h]
\centering
\includegraphics[height=\fgx]{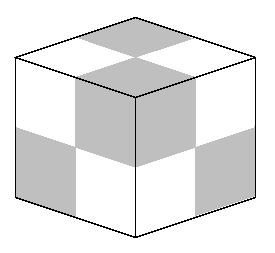}
\caption{A solid with the full tetrahedral cubic symmetry point group $T_d$ of hextetrahedral crystals.}
\end{figure}

Let us finally consider a crystalline insulator (or superconductor) with a full tetrahedral cubic symmetry point group $T_d$ which is one of the most intricate point group symmetries. As an abstract group, the symmetry group is given by the symmetric group \({G=S_4}\), and we mark the generators of its projective fermionic representation by \(\hat{c}_3,\hat{s}_4\) such that
\begin{equation}
\hat c_3^3=-1,\myspace \hat s_4^4=-1,\myspace (\hat{s}_4\hat c_3)^2=-1.
\end{equation}
The action of these generators on the 3D space is given by a threefold rotation for $\hat{c}_3$ and by a fourfold rotoreflection for $\hat{s}_4$ such that
\begin{equation}
\hat c_3^{\phantom{|}}\gamma_i\hat c_3^{-1}=\gamma_{i+1},\myspace \hat s_4^{\phantom{|}}\Big(\begin{smallmatrix}\gamma_1\\\gamma_2\\\gamma_3\end{smallmatrix}\Big)\hat s_4^{-1}=\Big(\begin{smallmatrix}-\gamma_2\\+\gamma_1\\-\gamma_3\end{smallmatrix}\Big).
\end{equation}
We wish to find the structure of the algebra \({B=\langle\Cl_{q+7,3},\Pi(S_4)\rangle}\) (and corresponding classifying space) generated by $\hat{c}_3,\hat{s}_4$ as well as by the Dirac gamma matrices, $\gamma_1,\gamma_2,\gamma_3$, and the other non spatial Clifford algebra generators \(\{\gamma\}\subset\{y\}\). In an analogous manner to Eq.~(\ref{rotrel}), we define new abstract generators \(c_3,s_4\) such that
\begin{equation}
c_3 = \tfrac{1+\gamma_1\gamma_2+\gamma_2\gamma_3+\gamma_3\gamma_1}{2}c_3,\myspace s_4 = e^{-\gamma_1\gamma_2\frac{\pi}{4}}\hat{s}_4\gamma_3.
\end{equation}
These new abstract generators satisfy
\begin{equation}\label{Tdrel}
\begin{gathered}
c_3^{\phantom{|}} y_i c_3^{-1}=y_i,\myspace s_4^{\phantom{|}} y_i s_4^{-1}=-y_i,\\ c_3^3=1,\myspace s_4^4=1,\myspace (s_4 c_3)^2=1.
\end{gathered}
\end{equation}
As above, these relations are exactly the relations of a $\ZZ_2$-graded tensor product \({B=\Cl_{q+7,3}\hat{\otimes}\RR[S_4]}\) where the ``even" part of $\RR[S_4]$, which commutes with the generators of $\Cl_{q+7,3}$, is generated by elements with $s_4$ occurring an even number of  times.
These even elements generate the subalgebra $\RR[A_4]$ of the alternating group \({A_4\triangleleft S_4}\). By knowing the grading of the groups \({S_4/A_4=\ZZ_2}\), we can find the grading of the group rings \({\RR[A_4]\hookrightarrow\RR[S_4]}\) as a direct sum of Clifford algebras
\begin{equation}
\begin{matrix}
\RR[A_4]&=&\RR\oplus\CC\oplus M_3(\RR)\\
\hookdownarrow&&\hookdownarrow\\
\RR[S_4]&=&\RR^\ptwo\oplus M_2(\RR)\oplus M_3(\RR)^\ptwo\\
&&\veq\\
&&\Cl_{0,1}\oplus\Cl_{0,2}\oplus(M_3(\RR)\otimes\Cl_{0,1}).
\end{matrix}
\end{equation}
We can thus find its graded tensor product with any Clifford algebra
\begin{equation}
\begin{aligned}
\Cl_{p,q}\hat\otimes\RR[S_4]&=\Cl_{p,q}\hat\otimes(\Cl_{0,1}\oplus\Cl_{0,2}\oplus(M_3(\RR)\otimes\Cl_{0,1}))\\ &=((\RR\oplus M_3(\RR))\otimes\Cl_{p,q+1})\oplus\Cl_{p,q+2}.
\end{aligned}
\end{equation}
And specifically for \({B=\Cl_{q+7,3}\hat{\otimes}\RR[S_4]}\) we find
\begin{multline}\label{resTd}
((\RR\oplus M_3(\RR))\otimes\Cl_{q+7,3+1})\oplus\Cl_{q+7,3+2}\\
\mapsto\sR_{q-4}^2\times\sR_{q-5},
\end{multline}
where we have once again used the mapping in Sec.~\ref{sec:ext} to determine the classifying space.

The classifying spaces of the complex AZ classes immediately follow by utilizing $\CC\otimes\Cl_{p,q}=\Cl_{p+q}$:
\begin{equation}\label{resTdc}
((\RR\oplus M_3(\RR))\otimes\Cl_{q+11})\oplus\Cl_{q+12}
\mapsto\sC_{q}^2\times\sC_{q+1}.
\end{equation}

The topological indices classifying materials with $T_d$ full tetrahedral symmetry, presented in Table~\ref{tab:exTd}, are the topological invariants of these classifying spaces, Eqs.~(\ref{resTd}),~(\ref{resTdc}).

The similar analyses of Appendix~\ref{app:pgf} are used to determine the classifying spaces in Table~\ref{tab:class} and the topological invariants in Tables~\ref{tab:p1}-\ref{tab:m1}.

\subsubsection*{Model Hamiltonians and Topological Invariants}
In order to construct a model Hamiltonian for these crystalline phases, one can use the model Hamiltonians of the tenfold-way classification with non-spatial symmetries~\cite{Schnyder2008Classification,ryu2010topological}, $\mathcal{H}_{q,d}^{\RR/\CC}$, for real/complex AZ class $q$ in $d$ spatial dimensions; see Sec.~\ref{rgtp}.

As a consequence of Eq.~(\ref{Tdrel}) one can write a model Hamiltonian, $\mathcal{H}_{q,3}^{\RR,T_d}$, in a block diagonal form, labelled by the irreducible representations of the subgroup $A_4$ generated by $c_3$ and $s_4^2$:
\begin{equation}
\mathcal{H}_{q,3}^{\RR,T_d}=\begin{pmatrix}
\big[\mathcal{H}_{q-1,3}^{\RR}\big]_{A} & 0 & 0\\
0 & \big[\mathcal{H}_{q-2,3}^{\RR}\big]_{E} & 0\\
0 & 0 & \big[\mathcal{H}_{q-1,3}^{\RR}\big]_{T}
\end{pmatrix}.
\end{equation}
Here, we use the Mulliken symbols~\cite{cotton2003chemical}, $A$, $E$, and $T$, to identify the irreducible representations corresponding to the subalgebra ${\RR[A_4]=\RR\oplus\CC\oplus M_3(\RR)}$. The complex AZ class yields a similar decomposition,
\begin{equation}
\mathcal{H}_{q,3}^{\CC,T_d}=\begin{pmatrix}
\big[\mathcal{H}_{q+1,3}^{\CC}\big]_{A} & 0 & 0\\
0 & \big[\mathcal{H}_{q,3}^{\CC}\big]_{E} & 0\\
0 & 0 & \big[\mathcal{H}_{q+1,3}^{\CC}\big]_{T}
\end{pmatrix}.
\end{equation}
The topological invariants of each block are the topological invariants of the corresponding Hamiltonians~\cite{Schnyder2008Classification,ryu2010topological,Shiozaki2014Topology,Chiu2016Classification}; see Table~\ref{tab:inv}.

In fact, the above examples in Secs.~\ref{sec:c3} and \ref{sec:s4} may also be recast into irreducible representation form by similarly identifying the subalgebras ${\RR[\ZZ_3]=\RR\oplus\CC}$ and ${\RR[\ZZ_2]=\RR\oplus\RR}$ with the Mulliken symbols $\{A,E\}$ and $\{A,B\}$, respectively. Moreover, one may use analogous irreducible representation analyses of the other point groups (see Sec.~\ref{sec:pgs} and Ref.~[\onlinecite{cotton2003chemical}]) in order to construct model Hamiltonians for all crystalline insulators and superconductors presented in Tables~\ref{tab:p1}-\ref{tab:m1}; see Sec.~\ref{rgtp}.

\section{Discussion}\label{sec:discussion}

In this paper, we have presented a complete classification of bulk topological invariants of crystalline topological insulators and superconductors in all AZ symmetry classes protected by all 32 point group symmetries of 3D crystals as well as by all 31 symmorphic layer group symmetries of 2D crystals. The majority of phases found by our classification paradigm are indeed novel crystalline topological insulators and superconductors. However, this classification is not exhaustive as crystals in nature may also have a nonsymmorphic magnetic space/layer group symmetry~\cite{graph}. In this discussion, we compare our results to those of previous works which have provided other classification schemes, emphasize the similarities and differences, and present some prospective ideas as to gaining a future exhaustive classification scheme of all topological phases of crystalline matter. Moreover, the experimental edge content signatures of many topological invariants are yet unclear, we thus also discuss the possible classification of edge content related to the invariants of crystalline topological phases.

\subsection{Magnetic Crystals}\label{sec:mag}

Although not directly calculated in this paper, it is relatively straightforward to generalize our results and classify topological phases protected by any of the 122 magnetic point group symmetries~\cite{graph} (also known as double point groups) of magnetic crystals~\cite{Shiozaki2014Topology,Schindler2018Higher,Sato2009Topological,Mong2010Antiferromagnetic,Mizushima2012Symmetry,mizushima2013topological,Fang2013Topological,liu2013antiferromagnetic,kotetes2013classification,Fang2014New,Zhang2015Topological,Watanabe2018Structure,lifshitz2005magnetic}. These crystals are invariant under elements of a group \({M=\langle N,(G\setminus N)1'\rangle}\) for some normal subgroup \({G/N=\ZZ_2}\) of the crystallographic point group \(G\). Here, we use \({1'^2=1}\) to be the symmetry action flipping the spin direction. Our methods can be directly extended to treat such cases; as an example we present the results for the simplest case of \(C_{2n}'\) symmetry in Table~\ref{tab:mag}; see Appendix~\ref{app:mag}. Order-two magnetic point groups~\cite{graph} such as \(C_2'\) were previously treated by Morimoto and Furusaki~[\onlinecite{Morimoto2013Topological}] and by Shiozaki and Sato [\onlinecite{Shiozaki2014Topology}], and our results are in complete agreement. The special case of \(C_4'\) was recently treated by Schindler \textit{et al.}~[\onlinecite{Schindler2018Higher}] where a \(\ZZ_2\) classification of the higher-order hinge states was found and is also in complete agreement with our results. We henceforth discuss such states in detail. Note, that a complete classification of the symmetry indicators of band structure topology of crystals with all 1651 magnetic space groups~\cite{graph} in AZ classes A and AI was recently carried out in Ref.~[\onlinecite{Watanabe2018Structure}].

\begin{table}[t]
\begin{equation*}
\begin{array}{l||c|c|c}
\mathrm{Sch\ddot{o}nflies} & C_2' & C_4' & C_6'
\\ \hline
\mathrm{HM} & 2' & 4' & 6'
\\ \hline\hline
T^2=-1 & \sR_{4-d} & \sR_{4-d}\times\sR_{-d} & \sR_{4-d}\times\sC_{d}
\\ \hline
d=3 &
\ZZ_2	& \ZZ_2 & \ZZ_2
\\
d=2 &
\ZZ_2	& \ZZ_2 & \ZZ_2\times\ZZ
\\ \hline\hline
T^2=+1 & \sR_{-d} & \sR_{-d}\times\sR_{4-d} & \sR_{-d}\times\sC_{d}
\\ \hline
d=3 &
0	& \ZZ_2 & 0
\\
d=2 &
0	& \ZZ_2 & \ZZ
\end{array}
\end{equation*}
\caption{Bulk topological invariants for magnetic point groups $C_{2n}'$ in 2 and 3 spatial dimensions.}
\label{tab:mag}
\end{table}

\subsection{Edge States and Higher Order Topological Insulators and Superconductors}

Until recently, bulk-boundary correspondence was considered a defining hallmark of topological phases~\cite{franz2013topological,Kane2005Topological,Kane2005Quantum,FuKaneMele2007Topological,Moore2007Topological,Hsieh2009Observation,Roy2009Topological,Fu2007Topological,Konig2007Quantum,xia2009observation,Chen2009Experimental}. Nevertheless, the recent discovery of higher-order topological insulators and superconductors drastically changed our understanding; some crystalline topological phases host no boundary modes~\cite{Hughes2011Inversion,Shiozaki2014Topology} while others host either boundary states~\cite{Shiozaki2015Z2,Khalaf2018Symmetry} or more exotic hinge or corner states~\cite{parameswaran2017topological,Benalcazar2017Quantized,Benalcazar2017Electric,Song2017d,Langbehn2017Reflection,Schindler2018Higher,schindler2018bismuth,xu2017topological,Shapourian2018Topological,lin2017topological,Ezawa2018Higher,Khalaf2018Higher,Geier2018Second,trifunovic2018higher,fang2017rotation}. A full treatment of the surface states of 3D crystalline topological insulators in AZ class AII was recently carried out by Khalaf \textit{et al.}~[\onlinecite{Khalaf2018Symmetry}] for all space group symmetries~\cite{graph}. Their analysis indicates that all surface states are in fact projections of the $\Gamma$-point Hamiltonian [see Eq.~(\ref{gammapt})], and indeed, when comparing with our results for all point group symmetries, we find the surface state indices to be subgroups of our bulk invariants. The classification of higher-order topological insulators and superconductors with order-two symmetries~\cite{graph} was recently accomplished by Geier \textit{et al.}~[\onlinecite{Geier2018Second}], Trifunovic and Brouwer~[\onlinecite{trifunovic2018higher}], and Khalaf~[\onlinecite{Khalaf2018Higher}]; we believe it should be possible to combine their methods with ours to achieve a full classification of higher-order topological invariants for all crystalline symmetries. Note, that parallel work in the complex AZ class A by Okuma \textit{et al.}~[\onlinecite{okuma2018topological}] achieved the classification of higher-order topological phases for all magnetic point group symmetries, and their bulk topological invariants in AZ class A are in complete agreement with our results for all point group symmetries.

\subsection{Full Brillouin zone structure, Symmetry indicators, and ``Weak" Topological Invariants}
Even without crystalline symmetry, the strong bulk invariant of the tenfold-way classification~\cite{kitaev2009periodic,schnyder2009classification,ryu2010topological,Hasan2010Colloquium,moore2010birth,franz2013topological,witten2015three} (see Table~\ref{tab:per}) is not the only possible topological invariant characterizing the material. Non-trivial topology may also occur along lower-dimensional surfaces(/curves) within the BZ torus $T^d$; these are known as ``weak" topological insulators and superconductors~\cite{FuKaneMele2007Topological,Moore2007Topological}. In fact, in AZ class AII, for example, the cellular (CW-complex) decomposition of the torus \({T^3=e^0\cup3e^1\cup3e^2\cup e^3}\) gives rise~\cite{kitaev2009periodic,Kennedy2015Homotopy} to the three ``weak" $\ZZ_2$ topological indices (see Table~\ref{tab:per})
\begin{equation}
\underbrace{\ZZ\ \times\ 0^3\ \times\ \ZZ_2^3}_{\mathrm{weak}}\ \times\underbrace{\ZZ_2}_{\mathrm{strong}}.
\end{equation}
The introduction of the crystalline structure complicates this simple relation between the strong bulk invariant and the weak ones~\cite{Varjas2017Space}.

Using the elementary band representations approach it is relatively easy to get the complete symmetry indicators for the full BZ torus~\cite{Dong2016Classification,po2017symmetry,bradlyn2017topological,Watanabe2018Structure,Bradlyn2018Band,song2018quantitative,Cano2018Building,Vergniory2017Graph,Ono2018Unified,vergniory2018high}. However, one still has to explicitly evaluate the Berry phases through various surfaces and curves (and Berry phases thereof) to find the different topological phases sharing an elementary band representation.

Using the K-theoretic approach~\cite{freed2013twisted,Shiozaki2017Topological,shiozaki2018atiyah} one studies the equivariant symmetry group action on the BZ torus ($G$-CW complex). This approach was successfully utilized by Shiozaki, Sato, and Gomi~[\onlinecite{Shiozaki2017Topological}] and yielded a complete classification for the wallpaper groups in the complex AZ classes (A and AIII). Indeed, when comparing with our results for the symmorphic layer groups~\cite{graph}, \(C_n,C_{nv}\), we find our bulk invariants to be subgroups of their full BZ torus K-groups.

\subsection{Defects and Higher Dimensional Systems}
Our paradigm is not restricted to point groups and can in fact be applied to classify crystalline topological phases in any spatial dimension, in particular, it can be applied to any of the 271 point groups (and 1202 magnetic point groups) of four-dimensional (4D) space. However, the immediate physical applicability of 4D crystals is less obvious than of 3D crystals, and so we leave this for prospecting future work.

A much more immediately relevant topic is that of crystals with defects. It has long been noticed~\cite{Teo2010Topological} that for the non-crystalline topological phases, all point, line, and surface defects may be easily incorporated into the classification schemes. This fact was explicitly shown to hold for crystalline materials with order-two~\cite{graph} symmetries~\cite{Shiozaki2014Topology} and had even also been formulated for the general crystalline case~\cite{Benalcazar2014Classification,Shiozaki2017Topological}.

Following Refs.~[\onlinecite{Shiozaki2017Topological,Teo2010Topological,stone2010symmetries}], let us observe a ${(\delta-1)}$-dimensional defect. It is surrounded by a sphere $S^D$ of codimension \({D=d-\delta}\); let us parametrize it by the spatial coordinates ${\mathbf{r}\in \RR^{D+1}}$ with \({||\mathbf{r}||=1}\). The crystalline symmetry $G$ leaves the defect invariant and hence acts separately on the momenta parallel to the defect \({\mathbf{k}_\parallel\in T^{d-D-1}}\) and on the momenta \({\mathbf{k}_\perp}\) conjugate to $\mathbf{r}$,
\begin{equation}
U_g\mathcal{H}(\mathbf{k},\mathbf{r})U_g^{-1}=\mathcal{H}(O_{g\parallel}\mathbf{k}_\parallel,O_{g\perp}\mathbf{k}_\perp,O_{g\perp}\mathbf{r}),
\end{equation}
where \({O_{g\parallel}\in\mathrm{O}(d-D-1)}\) and \({O_{g\perp}\in\mathrm{O}(D+1)}\) are orthogonal transformations; cf.~Eq.~(\ref{Og}). The algebraic structure may be restored by setting \({M(\mathbf{r})=\bm{\gamma}'\cdot\mathbf{r}}\) with \({\{\gamma'_i,\gamma'_j\}=-2\delta_{ij}}\) and \({\{\gamma'_i,\gamma_j\}=0}\) such that \({M^2=-1}\); cf.~Eq.~(\ref{gammapt}). Such analyses may be carried out in future works to classify the topological phases of defected crystals.

\subsection{``Fragile" Topological Phases}

It was recently noticed that some disconnected elementary band representations corresponding to non-trivial topological phases are trivializable by addition of trivial occupied bands~\cite{Po2018Fragile,bouhon2018wilson,bradlyn2018disconnected}, these were dubbed ``fragile" topological phases. It is often the case that in order to capture the topology of a macroscopically large number of bands, one introduces a stable-equivalence relation which disregards two phases as equivalent if they differ by addition of some trivial occupied or unoccupied bands. Such ``fragile" phases are missed by this stable-equivalence relation. The existence of similar phases was noticed even without crystalline symmetries, e.g., in Hopf topological insulators~\cite{Moore2008Topological,Deng2013Hopf} and Hopf topological superconductors~\cite{Kennedy2016Topological}. In order to generalize the results of our paper for such ``fragile" phases, one needs to avoid taking the stable limit in Eq.~(\ref{MapstoDef}) and count the connected components of the resulting topological space. However, more analysis is required as in the unstable case~\cite{kennedy2016bott}; not all bulk topologies are captured by a Dirac Hamiltonian~(\ref{gammapt}).

\section*{Acknowledgements}
All symmetry solids in this paper are taken with permission from Ashcroft and Mermin~\cite{ashcroft1976solid}. E.C.~acknowledges the help of his PhD advisor E.~Sela, and of the K-theory study group at Tel-Aviv University for providing the required background. A.C.~thanks Perimeter Institute for Theoretical Physics for the hospitality in the Summer of 2018, during which most of the work on this paper was carried out. The authors thank  B.~A.~Bernevig, T.~Neupert, I.~Le, R.~Lifshitz, R.~Ilan, A.~Bouhon, and J.~Hoeller for the fruitful discussions. 

\appendix

\section{Hierarchy of Symmetry Groups}\label{app:symmgraph}
\renewcommand{\thefigure}{A\arabic{figure}}
\setcounter{figure}{0}

\begin{figure}[t]
\centering
\includegraphics[width=\linewidth]{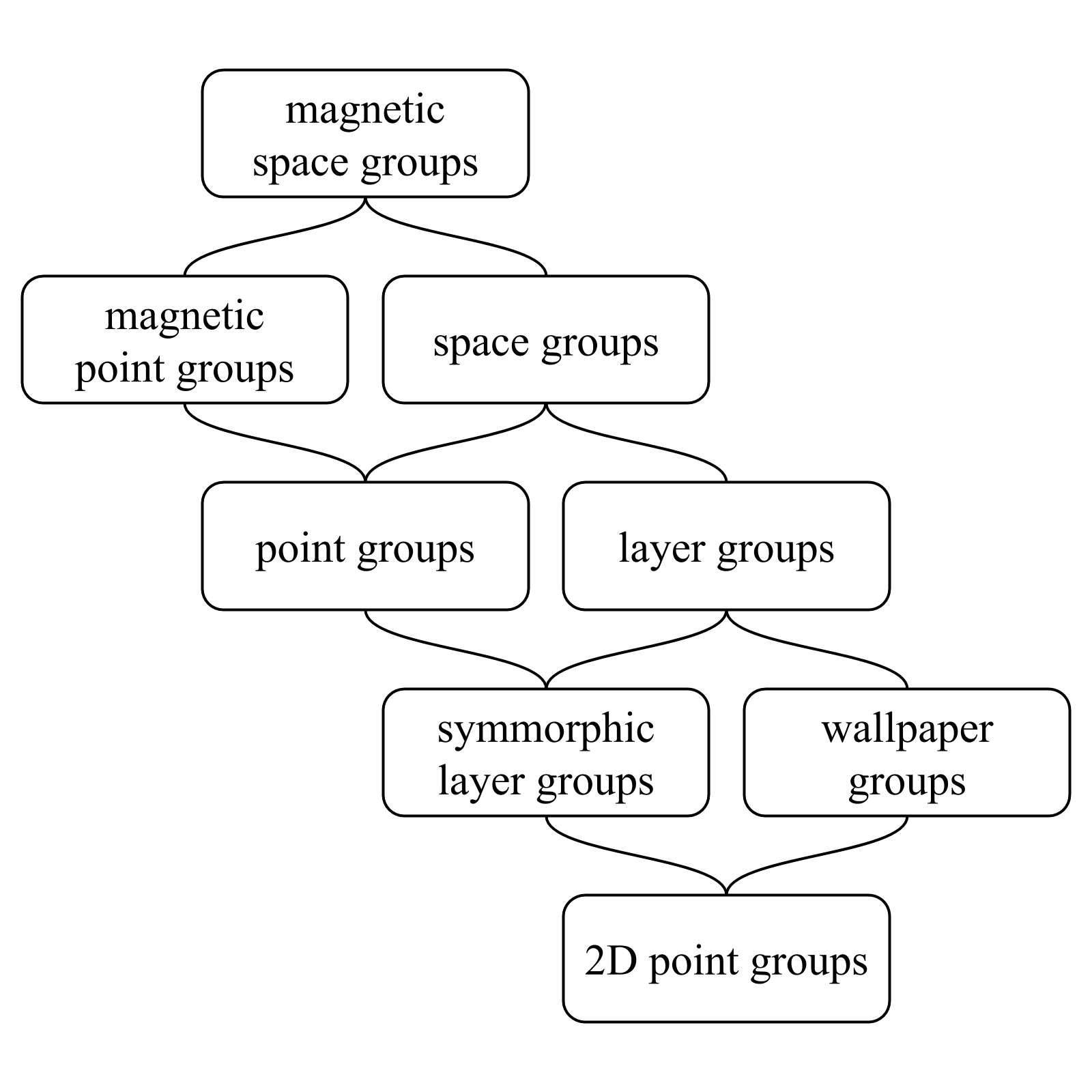}
\caption{Hierarchy of symmetry group classes.
}\label{fig:graph}
\end{figure}

In this appendix, we present the hierarchical structure of point group classes.

The hierarchical structure is presented as a graph in Fig.~\ref{fig:graph}. The intersection of every two adjacent symmetry group classes is given below them; e.g., the 2D point groups are all symmetry groups which are both wallpaper groups and symmorphic layer groups. Note, however, that there are some ``accidental" isomorphisms and so, for example, the distinct symmorphic layer groups $C_{1h}$ and $C_{1v}$ are isomorphic as point groups and sometimes denoted $C_s$.
Also note, that every class has a larger ``magnetic" variant (e.g., magnetic layer groups) and we omit them from the graph for simplicity.

In the paper we sometimes refer to ``order-two" symmetries. An order-two symmetry, in any symmetry group class, is a symmetry which is generated by a single element $g$ such that ${g^2=1}$. However, its projective representation may be non trivial ${\hat{g}^2=\pm1}$, and it may also be realized by antiunitary operators such as time-reversal (e.g., ${\hat{g}=\hat{c}_2T}$) in the magnetic classes. Examples of order-two symmetries include \(C_2,C_{1h},C_{1v},S_2,D_1\).

The constituents of actions of a symmetry group in any of the classes are as follows:

(i) 2D point groups - rotations and inversions (and their combined actions, i.e., reflections and rotoreflections) of a 2D space.

(ii) Wallpaper groups - rotations, inversions, and translations (and their combined actions, e.g., glides) of a 2D space.

(iii) Symmorphic layer groups - rotations and inversions of a 2D surface in a 3D space.

(iv) Layer groups - rotations, inversions and translations of a 2D surface in a 3D space.

(v) Point groups - rotations and inversions of a 3D space.

(vi) Space groups - rotations, inversions, and translations of a 3D space.

(vii) Magnetic point groups - point groups with time-reversal.

(viii) Magnetic space groups - space groups with time-reversal.

A symmetry group is dubbed nonsymmorphic if an element within it acts by a translation combined with any of the other constituents.

\section{From Graded Algebra Extensions to Classifying Spaces}\label{app:class-space}
\renewcommand{\theequation}{B\arabic{equation}}
\setcounter{equation}{0}

In this appendix, following Abramovici and Kalugin~[\onlinecite{abramovici2012clifford}], we give the algebraic derivation of the classifying spaces from algebra extensions.

We wish to study the extension of \(\ZZ_2\)-graded \(\RR\)-algebras,
\begin{equation}
B^0\hookrightarrow B=B^0\oplus B^1,
\end{equation}
graded by an element \(x_0\in B^1\).
Let us look at a \(B\)-module \(E\), it is also a \(B^0\)-module. In how many ways can one define a \(B\)-module structure on \(E\)? i.e., define a map \(\varphi\in\Hom_{B^0-\mathbf{alg}}(B,\End_{B^0}{E})\).
Such an action would act as \(\varphi(x_{0})\) on \(E\) as a \(B^0\)-module, however since \(E\) has a \(B\)-module structure than the most general action must be one may simply map \(E\) to the category of \(B\)-modules, act by \(x_{0}\), and map back, it is useful to write
\begin{gather}
\varphi(x_{0})=V^{-1}\alpha(x_{0})V,\\
\alpha\in\frac{\Hom_{B^0-\mathbf{alg}}(B,\End_{B^0}{E})}{V^{-1}\varphi V\simeq\varphi},\\
V\in\Aut_{B^0}{E}.
\end{gather}

What are the choices of \(\varphi\)? Let us focus on the case where \(B=\Cl_{p+1,q}\) and \(B^0=\Cl_{p,q}\).

If \(B\) is simple (Morita equivalent to \(\RR, \CC, \HH\)), then \(E=\RR^k\otimes_\RR\Lambda^2\) where \(\Lambda=\RR^{2^{p+q}}\) and \(\Lambda^2\) is a simple \(B\)-module; pick an action \(\alpha_0\in\Hom_{B^0-\mathbf{alg}}(B,\End_{B^0}{\Lambda})\). Moreover, as a \(B^0\)-module either \(E=(\RR^k\otimes_\RR\Lambda)\oplus(\RR^k\otimes_\RR\Lambda)\) or \(E=\RR^{2k}\otimes_\RR\Lambda\). Therefore, the only choice of \(\alpha\) is
\begin{align}
\alpha\in&\frac{\Hom_{B^0-\mathbf{alg}}(B,\End_{B^0}{E})}{V^{-1}\varphi V\simeq\varphi} \nonumber\\
&=\frac{(\alpha_0\circ(\Aut_{B^0}{B}))^{\oplus k}}{V^{-1}\varphi V\simeq\varphi}=[\alpha_0]^{\oplus k},
\end{align}
where the last equality follows from Skolem–-Noether.
On the other hand, this action is invariant under choices of \(V\in\Aut_{B^0}{E}\) that commute with \(\alpha_0(x_{0})\) which are just \(\Aut_B{E}\triangleleft\Aut_{B^0}{E}\) so
\begin{equation}
\varphi\in\frac{\Aut_{B^0}{E}}{\Aut_B{E}}.
\end{equation}
This space is homotopic to an appropriate symmetric space of \(\sR_1,\sR_2,\sR_3,\sR_5,\sR_6,\sR_7,\sC_1\); see Table~\ref{tab:per}.

If \(B=A\oplus A\) on the other hand (Morita equivalent to \(\RR\oplus\RR, \CC\oplus\CC, \HH\oplus\HH\)), then clearly \(E\simeq E_1\oplus E_2\) where \(E_{1,2}=\RR^{k_{1,2}}\otimes_\RR\Lambda\) and \(\Lambda=\RR^{2^{p+q}}\) is a simple \(A\)-module; pick actions \(\alpha_{k_1,k_2}\in\Hom_{B^0-\mathbf{alg}}(B,E_1\oplus E_2)\). One should thus consider all possible splittings \(E=E_1\oplus E_2\)
\begin{equation}
\alpha\in\frac{\Hom_{B^0-\mathbf{alg}}(B,\End_{B^0}{E})}{V^{-1}\varphi V\simeq\varphi}=\bigcup_{E_1\oplus E_2=E}[\alpha_{k_1,k_2}].
\end{equation}
The options for the action of \(x_{0}\) are thus
\begin{align}
\varphi\in&\bigcup_{E_1\oplus E_2=E}\frac{\Aut_{B^0}{E}}{\Aut_B(E_1\oplus E_2)} \nonumber\\
&=\bigcup_{E_1\oplus E_2=E}\frac{\Aut_{B^0}{E}}{\Aut_{B^0}{E_1}\times\Aut_{B^0}{E_2}}.
\end{align}
This space is homotopic to an appropriate symmetric space of \(\sR_0,\sR_4,\sC_0\); see Table~\ref{tab:per}.

\section{Clifford Algebras}\label{app:cliff}
\renewcommand{\thetable}{C-\Roman{table}}
\setcounter{table}{0}

In this appendix we give some of some complex and real Clifford algebras, they are presented in Table~\ref{tab:clifC} and Table~\ref{tab:clifR}.

\begin{table}[h]
\begin{equation*}
\begin{array}{c||ccc}	
q & 0 & 1 & 2
\\\hline\hline
\Cl_{q} & \CC & \CC^\ptwo & M_2(\CC)
\end{array}
\end{equation*}
\caption{Complex Clifford algebras, $\Cl_{q}$.}\label{tab:clifC}
\end{table}

\begin{table*}[t]
\begin{equation*}
\begin{array}{c||ccccccccc}	
p\setminus q & 0 & 1 & 2 & 3 & 4 & 5 & 6 & 7 & 8
\\\hline
\mathrm{``-1"}&  & \RR & \CC & \HH & \HH^\ptwo & M_2(\HH) & M_4(\CC) & M_8(\RR) & M_8(\RR)^\ptwo
\\\hline\hline
0 & \RR & \RR^\ptwo & M_2(\RR) & M_2(\CC) & M_2(\HH) & M_2(\HH)^\ptwo & M_4(\HH) & M_8(\CC) & M_{16}(\RR)
\\
1 & \CC & M_2(\RR) & M_2(\RR)^\ptwo & M_4(\RR) & M_4(\CC) & M_4(\HH) & M_4(\HH)^\ptwo & M_8(\HH) & M_{16}(\CC)
\\
2 & \HH & M_2(\CC) & M_4(\RR) & M_4(\RR)^\ptwo & M_8(\RR) & M_8(\CC) & M_8(\HH) & M_8(\HH)^\ptwo & M_{16}(\HH)
\\
3 & \HH^\ptwo & M_2(\HH) & M_4(\CC) & M_8(\RR) & M_8(\RR)^\ptwo & M_{16}(\RR) & M_{16}(\CC) & M_{16}(\HH) & M_{16}(\HH)^\ptwo
\\
4 & M_2(\HH) & M_2(\HH)^\ptwo & M_4(\HH) & M_8(\CC) & M_{16}(\RR) & M_{16}(\RR)^\ptwo & M_{32}(\RR) & M_{32}(\CC) & M_{32}(\HH)
\\
5 & M_4(\CC) & M_4(\HH) & M_4(\HH)^\ptwo & M_8(\HH) & M_{16}(\CC) & M_{32}(\RR) & M_{32}(\RR)^\ptwo & M_{64}(\RR) & M_{64}(\CC)
\\
6 & M_8(\RR) & M_8(\CC) & M_8(\HH) & M_8(\HH)^\ptwo & M_{16}(\HH) & M_{32}(\CC) & M_{64}(\RR) & M_{64}(\RR)^\ptwo & M_{128}(\RR)
\\
7 & M_8(\RR)^\ptwo & M_{16}(\RR) & M_{16}(\CC) & M_{16}(\HH) & M_{16}(\HH)^\ptwo & M_{32}(\HH) & M_{64}(\CC) & M_{128}(\RR) & M_{128}(\RR)^\ptwo
\\
8 & M_{16}(\RR) & M_{16}(\RR)^\ptwo & M_{32}(\RR) & M_{32}(\CC) & M_{32}(\HH) & M_{32}(\HH)^\ptwo & M_{64}(\HH) & M_{128}(\CC) & M_{256}(\RR)
\end{array}
\end{equation*}
\caption{Real Clifford algebras, $\Cl_{p,q}$. The even part of every algebra $\Cl_{p,q}$ is isomorphic to the algebra $\Cl_{p-1,q}$ which is one row above it.}\label{tab:clifR}
\end{table*}

\section{Point Group Symmetry Classification}\label{app:pgf}
\renewcommand{\theequation}{D\arabic{equation}}
\setcounter{equation}{0}
\renewcommand{\thefigure}{D\arabic{figure}}
\setcounter{figure}{0}

In this appendix we derive the bulk invariants of all 32 point group symmetries of 3D crystals. This is done using the techniques demonstrated in Sec.~\ref{sec:exam}.

\subsection{Rotational Symmetry \(C_n\)}
\begin{figure}[h]
\centering
\includegraphics[height=\fg]{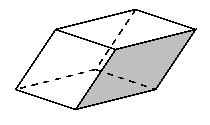}~\includegraphics[height=\fg]{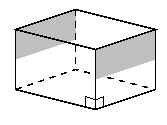}~\includegraphics[height=\fg]{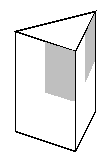}~\includegraphics[height=\fg]{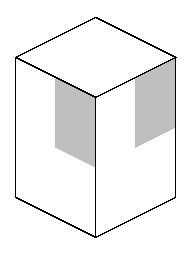}~\includegraphics[height=\fg]{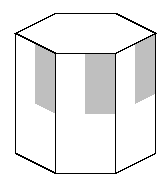}
\caption{The \(C_n\) rotational symmetry point groups: \(C_1,C_2,C_3,C_4,C_6\).}
\end{figure}

The symmetry group is \({G=\ZZ_n}\) with generator \(\hat{c}\) such that
\begin{equation}
\hat c^n=-1.
\end{equation}
It acts on the 3D space as
\begin{equation}
\hat c(\begin{smallmatrix}\gamma_1\\\gamma_2\end{smallmatrix})\hat c^{-1}=(\begin{smallmatrix}\cos\frac{2\pi}{n}&\sin\frac{2\pi}{n}\\-\sin\frac{2\pi}{n}&\cos\frac{2\pi}{n}\end{smallmatrix})(\begin{smallmatrix}\gamma_1\\\gamma_2\end{smallmatrix}).
\end{equation}
Notice that
\begin{equation}
\begin{aligned}
e^{-\gamma_1\gamma_2\frac{2\pi}{2n}}(\begin{smallmatrix}\gamma_1\\\gamma_2\end{smallmatrix})e^{\gamma_1\gamma_2\frac{2\pi}{2n}}
&=e^{-\gamma_1\gamma_2\frac{2\pi}{n}}(\begin{smallmatrix}\gamma_1\\\gamma_2\end{smallmatrix}) \\ &=(\begin{smallmatrix}\cos\frac{2\pi}{n}&\sin\frac{2\pi}{n}\\-\sin\frac{2\pi}{n}&\cos\frac{2\pi}{n}\end{smallmatrix})(\begin{smallmatrix}\gamma_1\\\gamma_2\end{smallmatrix}).
\end{aligned}
\end{equation}
We set
\begin{equation}
c = e^{\gamma_1\gamma_2\frac{2\pi}{2n}}\hat c,
\end{equation}
which satisfies
\begin{equation}
\myspace c y_i c^{-1}=y_i,\myspace c^n=1.
\end{equation}
We thus get \({B=\RR[\ZZ_n]\otimes\Cl_{q+7,3}}\) and hence
\begin{equation}
(\RR^{\oplus r(n)}\oplus\CC^{\oplus c(n)})\otimes\Cl_{q+7,3}\mapsto\sR_{q-3}^{r(n)}\times\sC_{q+1}^{c(n)}.
\end{equation}
Here, we have used the mapping in Sec.~\ref{sec:ext} to determine the classifying spaces; we employ this mapping throughout this section in treatment of all point group symmetries.

\subsection{Rotoreflection Symmetry \(S_{2n}\)}
\begin{figure}[h]
\centering
\includegraphics[height=\fg]{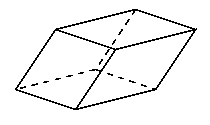}~\includegraphics[height=\fg]{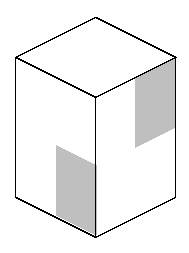}~\includegraphics[height=\fg]{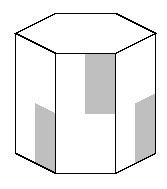}
\caption{The \(S_{2n}\) rotoreflection symmetry point groups: \(S_2,S_4,S_6\).}
\end{figure}

The symmetry group is \({G=\ZZ_{2n}}\) with generator \(\hat{s}\) such that
\begin{equation}
\hat s^{2n}=-(-1)^n.
\end{equation}
It acts on the 3D space as
\begin{equation}
\hat s(\begin{smallmatrix}\gamma_1\\\gamma_2\end{smallmatrix})\hat s^{-1}=(\begin{smallmatrix}\cos\frac{2\pi}{2n}&\sin\frac{2\pi}{2n}\\-\sin\frac{2\pi}{2n}&\cos\frac{2\pi}{2n}\end{smallmatrix})(\begin{smallmatrix}\gamma_1\\\gamma_2\end{smallmatrix}),\myspace \hat s\gamma_3 \hat s^{-1}=-\gamma_3.
\end{equation}
We set
\begin{equation}
s = e^{\gamma_1\gamma_2\frac{2\pi}{4n}}\hat s\gamma_3,
\end{equation}
which satisfies
\begin{equation}
\myspace s y_i s^{-1}= -y_i,\myspace s^{2n}=1.
\end{equation}
Here we used the equality \(s^{2n}=(-e^{-\gamma_1\gamma_2\frac{2\pi}{2n}}\hat s^2)^n=1\).

For \(n=1\) we add \(s\) as a generator such that \(\RR\langle s\rangle=\Cl_{0,1}\) and \({\Cl_{p,q}\hat{\otimes}\Cl_{0,1}=\Cl_{p,q+1}}\), and get
\begin{equation}
\Cl_{q+7,3+1}\mapsto\sR_{q-4}.
\end{equation}

For \(n=2\) we have a central element \({t=s^2}\) satisfying ${t^2=1}$. Our algebra decomposes accordingly as a direct sum of two algebras such that in one of them \(s^2=t=1\) and in the other \(s^2=t=-1\). Therefore, we get
\begin{equation}
\Cl_{q+7,3+1}\oplus\Cl_{q+7+1,3}\mapsto\sR_{q-4}\times\sR_{q-2}.
\end{equation}
This can also be done by considering \({\ZZ_2\triangleleft\ZZ_4}\) such that the graded structure is
\begin{equation}
\begin{matrix}
\RR[\ZZ_2]&=&\RR\oplus\RR\\
\hookdownarrow&&\hookdownarrow&&\\
\RR[\ZZ_4]&=&\RR^\ptwo\oplus\CC&=&\Cl_{0,1}\oplus\Cl_{1,0},
\end{matrix}
\end{equation}
\begin{equation}
\begin{aligned}
\Cl_{p,q}\hat{\otimes}\RR[\ZZ_4]&=\Cl_{p,q}\hat{\otimes}(\Cl_{0,1}\oplus\Cl_{1,0})\\&=\Cl_{p,q+1}\oplus\Cl_{p+1,q}.
\end{aligned}
\end{equation}
Such analyses would come in handy in other more complicated cases.

For \(n=3\) we use \(\ZZ_6=\ZZ_2\times\ZZ_3\) and split our algebra to \({\Cl_{p,q}\hat\otimes\RR\langle s\rangle=(\Cl_{p,q}\hat\otimes\RR\langle s^3\rangle)\otimes\RR\langle s^2\rangle}\) and get
\begin{equation}
(\RR\oplus\CC)\otimes\Cl_{q+7,3+1}\mapsto\sR_{q-4}\times\sC_{q}.
\end{equation}

\subsection{Dipyramidal Symmetry \(C_{nh}\)}
\begin{figure}[h]
\centering
\includegraphics[height=\fg]{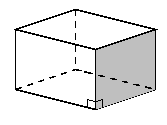}~\includegraphics[height=\fg]{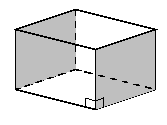}~\includegraphics[height=\fg]{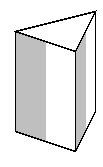}~\includegraphics[height=\fg]{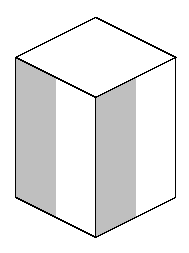}~\includegraphics[height=\fg]{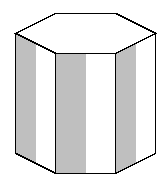}
\caption{The \(C_{nh}\) point groups: \(C_{1h},C_{2h},C_{3h},C_{4h},C_{6h}\).}
\end{figure}

The symmetry group is \({G=\ZZ_n\times\ZZ_2}\) with generators \(\hat{c},\hat{\sigma}_h\) such that
\begin{equation}
\hat c^n=-1,\myspace \hat{\sigma}_h^2=-1.
\end{equation}
It acts on the 3D space as
\begin{equation}
\hat c(\begin{smallmatrix}\gamma_1\\\gamma_2\end{smallmatrix})\hat c^{-1}=(\begin{smallmatrix}\cos\frac{2\pi}{n}&\sin\frac{2\pi}{n}\\-\sin\frac{2\pi}{n}&\cos\frac{2\pi}{n}\end{smallmatrix})(\begin{smallmatrix}\gamma_1\\\gamma_2\end{smallmatrix}),\myspace \hat{\sigma}_h^{\phantom{|}}\gamma_3 \hat{\sigma}_h^{-1}=-\gamma_3.
\end{equation}
We set
\begin{equation}
c = e^{\gamma_1\gamma_2\frac{2\pi}{2n}}\hat c,\myspace \sigma_h = \hat{\sigma}_h\gamma_3,
\end{equation}
which satisfy
\begin{equation}
\begin{gathered}
c y_i c^{-1}=y_i,\myspace \sigma_h^{\phantom{|}} y_i \sigma_h^{-1}=-y_i,\myspace c^n=1,\myspace \sigma_h^2=1.
\end{gathered}
\end{equation}
We add \(\sigma_h\) as a generator and get \({\RR[\ZZ_n]\otimes(\Cl_{p,q}\hat\otimes\RR[\ZZ_2])}\) and hence
\begin{equation}
(\RR^{\oplus r(n)}\oplus\CC^{\oplus c(n)})\otimes\Cl_{q+7,3+1}\mapsto\sR_{q-4}^{r(n)}\times\sC_{q}^{c(n)}.
\end{equation}

\subsection{Pyramidal Symmetry \(C_{nv}\)}
\begin{figure}[h]
\centering
\includegraphics[height=\fg]{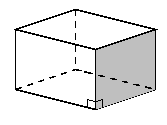}~\includegraphics[height=\fg]{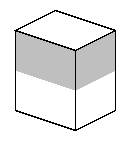}~\includegraphics[height=\fg]{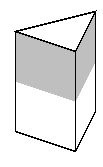}~\includegraphics[height=\fg]{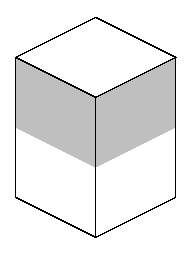}~\includegraphics[height=\fg]{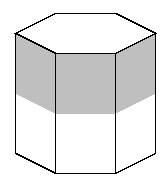}
\caption{The \(C_{nv}\) pyramidal symmetry point groups: \(C_{1v},C_{2v},C_{3v},C_{4v},C_{6v}\).}
\end{figure}

The symmetry group is \({G=\Dih_n}\) with generators \(\hat{c},\hat{\sigma}_v\) such that
\begin{equation}
\hat c^n=-1,\myspace \hat{\sigma}_v^2=-1,\myspace (\hat c\hat{\sigma}_v)^2=-1.
\end{equation}
It acts on the 3D space as
\begin{equation}
\hat c(\begin{smallmatrix}\gamma_1\\\gamma_2\end{smallmatrix})\hat c^{-1}=(\begin{smallmatrix}\cos\frac{2\pi}{n}&\sin\frac{2\pi}{n}\\-\sin\frac{2\pi}{n}&\cos\frac{2\pi}{n}\end{smallmatrix})(\begin{smallmatrix}\gamma_1\\\gamma_2\end{smallmatrix}),\myspace \hat{\sigma}_v^{\phantom{|}}\gamma_1 \hat{\sigma}_v^{-1}=-\gamma_1.
\end{equation}
We set
\begin{equation}
c = e^{\gamma_1\gamma_2\frac{2\pi}{2n}}\hat c,\myspace \sigma_v = \hat{\sigma}_v\gamma_1,
\end{equation}
which satisfy
\begin{equation}
\begin{gathered}
c y_i c^{-1}=y_i,\myspace \sigma_v^{\phantom{|}} y_i \sigma_v^{-1}=-y_i,\\ c^n=1,\myspace \sigma_v^2=1,\myspace (c\sigma_v)^2=1.
\end{gathered}
\end{equation}
Here we used the equality
\begin{equation}
\begin{aligned}
c\sigma_v &= e^{\gamma_1\gamma_2\frac{2\pi}{2n}}\hat c\hat\sigma_v\gamma_1 = \hat c\hat\sigma_v\gamma_1 e^{\gamma_1\gamma_2\frac{2\pi}{2n}} \\ &=  \hat\sigma_v e^{\gamma_1\gamma_2\frac{2\pi}{n}}\gamma_1 \hat c^{-1} e^{\gamma_1\gamma_2\frac{2\pi}{2n}} \\ &= \hat\sigma_v\gamma_1 \hat c^{-1} e^{-\gamma_1\gamma_2\frac{2\pi}{2n}} = \sigma_v c^{-1}.
\end{aligned}
\end{equation}
We find that the ring of \({\ZZ_n\triangleleft\Dih_n}\) commutes with \(\Cl_{p,q}\). Let us look at the algebra structure
\begin{equation}
\begin{matrix}
\RR[\ZZ_n]&=&\RR^{\oplus r(n)}\oplus\CC^{\oplus c(n)}\\
\hookdownarrow&&\hookdownarrow\\
\RR[\Dih_n]&=&(\RR^\ptwo)^{\oplus r(n)}\oplus M_2(\RR)^{\oplus c(n)}\\
&&\veq\\
&&\Cl_{0,1}^{\oplus r(n)}\oplus \Cl_{0,2}^{\oplus c(n)},
\end{matrix}
\end{equation}
\begin{equation}
\begin{aligned}
\Cl_{p,q}\hat\otimes\RR[\Dih_n]&=\Cl_{p,q}\hat\otimes(\Cl_{0,1}^{\oplus r(n)}\oplus \Cl_{0,2}^{\oplus c(n)})\\ &=\Cl_{p,q+1}^{\oplus r(n)}\oplus\Cl_{p,q+2}^{\oplus c(n)}.
\end{aligned}
\end{equation}
We thus get
\begin{equation}
\Cl_{q+7,3+1}^{\oplus r(n)}\oplus\Cl_{q+7,3+2}^{\oplus c(n)}\mapsto\sR_{q-4}^{r(n)}\times\sR_{q-5}^{c(n)}.
\end{equation}

\subsection{Dihedral Symmetry \(D_n\)}
\begin{figure}[h]
\centering
\includegraphics[height=\fg]{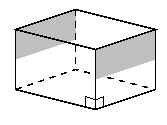}~\includegraphics[height=\fg]{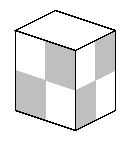}~\includegraphics[height=\fg]{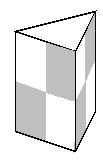}~\includegraphics[height=\fg]{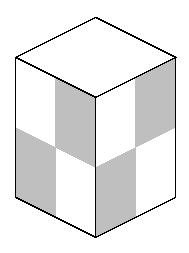}~\includegraphics[height=\fg]{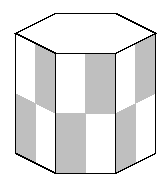}
\caption{The \(D_{n}\) dihedral symmetry point groups: \(D_{1},D_{2},D_{3},D_{4},D_{6}\).}
\end{figure}

The symmetry group is \({G=\Dih_n}\) with generators \(\hat{c}_n,\hat{c}_2\) such that
\begin{equation}
\hat c_n^n=-1,\myspace \hat c_2^2=-1,\myspace (\hat c_n\hat{c}_2)^2=-1.
\end{equation}
It acts on the 3D space as
\begin{equation}
\hat c_n^{\phantom{|}}(\begin{smallmatrix}\gamma_1\\\gamma_2\end{smallmatrix})\hat c_n^{-1}=(\begin{smallmatrix}\cos\frac{2\pi}{n}&\sin\frac{2\pi}{n}\\-\sin\frac{2\pi}{n}&\cos\frac{2\pi}{n}\end{smallmatrix})(\begin{smallmatrix}\gamma_1\\\gamma_2\end{smallmatrix}),\myspace \hat{c}_2^{\phantom{|}}\gamma_{2,3} \hat{c}_2^{-1}=-\gamma_{2,3}.
\end{equation}
We set
\begin{equation}
c_n = e^{\gamma_1\gamma_2\frac{2\pi}{2n}}\hat c_n,\myspace c_2 = \hat{c}_2\gamma_2\gamma_3,
\end{equation}
which satisfy
\begin{equation}
\begin{gathered}
c_n^{\phantom{|}} y_i c_n^{-1}=y_i,\myspace c_2^{\phantom{|}} y_i c_2^{-1}=y_i,\\ c_n^n=1,\myspace c_2^2=1,\myspace (c_n c_2)^2=1.
\end{gathered}
\end{equation}
We thus get \({B=\RR[\Dih_n]\otimes\Cl_{q+7,3}}\) and hence
\begin{equation}
((\RR^\ptwo)^{\oplus r(n)}\oplus M_2(\RR)^{\oplus c(n)})\otimes\Cl_{q+7,3}\mapsto\sR_{q-3}^{2r(n)+c(n)}.
\end{equation}

\subsection{Antiprismatic Symmetry \(D_{nd}\)}
\begin{figure}[h]
\centering
\includegraphics[height=\fg]{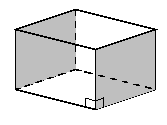}~\includegraphics[height=\fg]{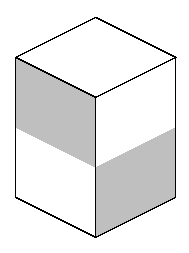}~\includegraphics[height=\fg]{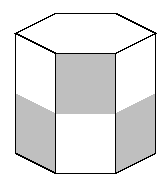}
\caption{The \(D_{nd}\) antiprismatic symmetry point groups: \(D_{1d},D_{2d},D_{3d}\).}
\end{figure}

The symmetry group is \({G=\Dih_{2n}}\) with generators \(\hat{s},\hat{\sigma}_v\) such that
\begin{equation}
\hat s^{2n}=-(-1)^n,\myspace \hat{\sigma}_v^2=-1,\myspace (\hat s\hat{\sigma}_v)^2=-1.
\end{equation}
It acts on the 3D space as
\begin{equation}
\begin{gathered}
\hat s(\begin{smallmatrix}\gamma_1\\\gamma_2\end{smallmatrix})\hat s^{-1}=(\begin{smallmatrix}\cos\frac{2\pi}{2n}&\sin\frac{2\pi}{2n}\\-\sin\frac{2\pi}{2n}&\cos\frac{2\pi}{2n}\end{smallmatrix})(\begin{smallmatrix}\gamma_1\\\gamma_2\end{smallmatrix}),\\ \hat s\gamma_3 \hat s^{-1}=-\gamma_3,\myspace \hat{\sigma}_v^{\phantom{|}}\gamma_1 \hat{\sigma}_v^{-1}=-\gamma_1.
\end{gathered}
\end{equation}
We set
\begin{equation}
s = e^{\gamma_1\gamma_2\frac{2\pi}{2n}}\hat s\gamma_3,\myspace \sigma_v = \hat{\sigma}_v\gamma_1,
\end{equation}
which satisfy
\begin{equation}
\begin{gathered}
s y_i s^{-1}=-y_i,\myspace \sigma_v^{\phantom{|}} y_i \sigma_v^{-1}=-y_i,\\ s^{2n}=1,\myspace \sigma_v^2=1,\myspace (s\sigma_v)^2=1.
\end{gathered}
\end{equation}
Here we used the equality
\begin{equation}
\begin{aligned}
s\sigma_v &= e^{\gamma_1\gamma_2\frac{2\pi}{4n}}\hat s\gamma_3\hat\sigma_v\gamma_1 = \hat s\gamma_3\hat\sigma_v\gamma_1 e^{\gamma_1\gamma_2\frac{2\pi}{4n}} \\ &=\hat\sigma_ve^{\gamma_1\gamma_2\frac{2\pi}{2n}}\gamma_1 \gamma_3\hat s^{-1} e^{\gamma_1\gamma_2\frac{2\pi}{4n}} \\ &= \hat\sigma_v\gamma_1 \gamma_3 \hat s^{-1} e^{-\gamma_1\gamma_2\frac{2\pi}{4n}} = \sigma_v s^{-1}.
\end{aligned}
\end{equation}

For \({n=1}\) we have \({D_{1d}=C_{2h}}\) by setting \({c = s\sigma_v},{\sigma_h = \sigma_v}\).

For \({n=2}\) we have a central element \({t = s^2}\) with ${t^2=1}$, and our algebra decomposes accordingly as a direct sum of simple algebras such that in the first component (isomorphic to  \(D_{1d}\)) we have \(t=-1\),  and in the second component (in which \({s^2=-1},{s\sigma_v=-\sigma_v s}\)) we have \(t=-1\). Therefore, we get
\begin{equation}
(\RR^\ptwo\otimes\Cl_{q+7,3+1})\oplus\Cl_{q+7+1,3+1} \mapsto\sR_{q-4}^2\times\sR_{q-3}.
\end{equation}
This can also be done by considering \({\Dih_2\triangleleft\Dih_4}\) such that the graded structure is
\begin{equation}
\begin{matrix}
\RR[\Dih_2]&=&\RR^\ptwo\oplus\RR^\ptwo&&\\
\hookdownarrow&&\hookdownarrow&&\\
\RR[\Dih_4]&=&(\RR^\ptwo)^\ptwo\oplus M_2(\RR)&=&\Cl_{0,1}^\ptwo\oplus\Cl_{1,1},
\end{matrix}
\end{equation}
\begin{equation}
\begin{aligned}
\Cl_{p,q}\hat{\otimes}\RR[\ZZ_4]&=\Cl_{p,q}\hat{\otimes}(\Cl_{0,1}^\ptwo\oplus\Cl_{1,1})\\&=\Cl_{p,q+1}^\ptwo\oplus\Cl_{p+1,q+1}.
\end{aligned}
\end{equation}

For \({n=3}\) we use \(\Dih_6=\ZZ_2\times\Dih_3\) and split our algebra to \({\Cl_{p,q}\hat\otimes\RR\langle s,\sigma_v\rangle=(\Cl_{p,q}\hat\otimes\RR\langle s^3\rangle)\otimes\RR\langle s^2,s\sigma_v\rangle}\) and get
\begin{equation}
(\RR^\ptwo\oplus M_2(\RR))\otimes\Cl_{q+7,3+1}\mapsto\sR_{q-4}^3.
\end{equation}

\subsection{Prismatic Symmetry \(D_{nh}\)}
\begin{figure}[h]
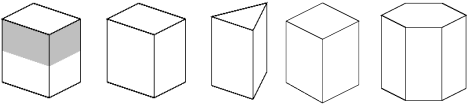

\centering
\includegraphics[height=\fg]{C2v_3D.png}~\includegraphics[height=\fg]{D2h_3D.png}~\includegraphics[height=\fg]{D3h_3D.png}~\includegraphics[height=\fg]{D4h_3D.png}~\includegraphics[height=\fg]{D6h_3D.png}
\caption{The \(D_{nh}\) prismatic symmetry point groups: \(D_{1h},D_{2h},D_{3h},D_{4h},D_{6h}\).}
\end{figure}

The symmetry group is \({G=\Dih_{n}\times\ZZ_2}\) with generators \(\hat{c}_n,\hat{c}_2,\hat{\sigma}_h\) such that
\begin{equation}
\begin{gathered}
\hat c_n^{n}=-1,\myspace \hat{c}_2^2=-1,\myspace \hat{\sigma}_h^2=-1,\\ (\hat c_n\hat{c}_2)^2=-1,\myspace (\hat c_2\sigma_h)^2=-1.
\end{gathered}
\end{equation}
It acts on the 3D space as 
\begin{equation}
\begin{gathered}
\hat c_n^{\phantom{|}}(\begin{smallmatrix}\gamma_1\\\gamma_2\end{smallmatrix})\hat c_n^{-1}=(\begin{smallmatrix}\cos\frac{2\pi}{n}&\sin\frac{2\pi}{n}\\-\sin\frac{2\pi}{n}&\cos\frac{2\pi}{n}\end{smallmatrix})(\begin{smallmatrix}\gamma_1\\\gamma_2\end{smallmatrix}),\\ \hat{c}_2^{\phantom{|}}\gamma_{2,3} \hat{c}_2^{-1}=-\gamma_{2,3},\myspace \hat{\sigma}_h^{\phantom{|}}\gamma_3 \hat{\sigma}_h^{-1}=-\gamma_3.
\end{gathered}
\end{equation}
We set
\begin{equation}
c_n = e^{\gamma_1\gamma_2\frac{2\pi}{2n}}\hat c_n,\myspace c_2 = \hat{c}_2\gamma_2\gamma_3,\myspace \sigma_h = \hat{\sigma}_h\gamma_3,
\end{equation}
which satisfy
\begin{equation}
\begin{gathered}
c_n^{\phantom{|}} y_i c_n^{-1}=y_i,\myspace c_2^{\phantom{|}} y_i c_2^{-1}=y_i,\myspace \sigma_h^{\phantom{|}} y_i \sigma_h^{-1}=-y_i,\\
c_n^{n}=1,\myspace c_2^2=1,\myspace \sigma_h^2=1,\myspace (c_nc_2)^2=1,\myspace (c_2\sigma_h)^2=1.
\end{gathered}
\end{equation}
We add \(\sigma_h\) as a generator and get \({\RR[\Dih_n]\otimes(\Cl_{p,q}\hat{\otimes}\RR[\ZZ_2])}\) and hence
\begin{equation}
((\RR^\ptwo)^{\oplus r(n)}\oplus M_2(\RR)^{\oplus c(n)})\otimes\Cl_{q+7,3+1} \mapsto\sR_{q-4}^{2r(n)+c(n)}.
\end{equation}

\begin{figure}[h]
\centering
\includegraphics[height=\fg]{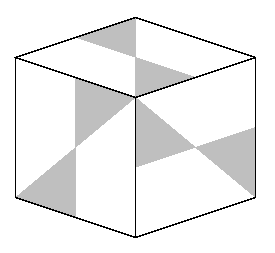}~\includegraphics[height=\fg]{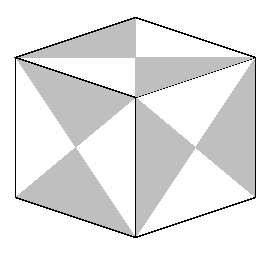}~\includegraphics[height=\fg]{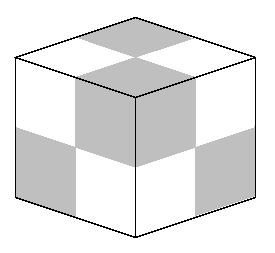}~\includegraphics[height=\fg]{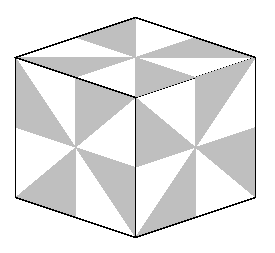}~\includegraphics[height=\fg]{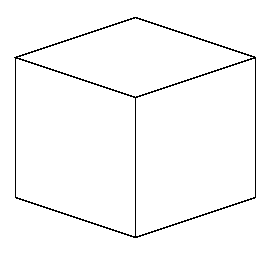}
\caption{The cubic point groups: \(T,T_h,T_d,O,O_h\).}
\end{figure}

\subsection{Chiral Tetrahedral Symmetry \(T\)}
The symmetry group is \({G=A_4}\) with generators \(\hat{c}_3,\hat{c}_2\) such that
\begin{equation}
\hat c_3^3=-1,\myspace \hat c_2^2=-1,\myspace (\hat{c}_2\hat c_3)^3=-1.
\end{equation}
It acts on the 3D space as
\begin{equation}
\hat c_3^{\phantom{|}}\gamma_i\hat c_3^{-1}=\gamma_{i+1},\myspace \hat{c}_2^{\phantom{|}}\gamma_{1,2} \hat{c}_2^{-1}=-\gamma_{1,2}.
\end{equation}
We set
\begin{equation}
c_3 = \tfrac{1+\gamma_1\gamma_2+\gamma_2\gamma_3+\gamma_3\gamma_1}{2}c_3,\myspace c_2 = -\hat{c}_2\gamma_1\gamma_2,
\end{equation}
which satisfy
\begin{equation}
\begin{gathered}
c_3^{\phantom{|}} y_i c_3^{-1}=y_i,\myspace c_2^{\phantom{|}} y_i c_2^{-1}=y_i,\\ c_3^3=1,\myspace c_2^2=1,\myspace (c_2 c_3)^3=1.
\end{gathered}
\end{equation}
Here we used the equality
\begin{equation}
\begin{aligned}
(c_2 c_3)^3 &= (\gamma_1\gamma_2 \hat c_2 \hat c_3)^3\\&=(\gamma_1\gamma_2)(-\gamma_2\gamma_3)(+\gamma_3\gamma_1)(\hat c_2 \hat c_3)^3=1.
\end{aligned}
\end{equation}
We thus get \({B=\RR[A_4]\otimes\Cl_{q+7,3}}\) and hence
\begin{equation}
(\RR\oplus\CC\oplus M_3(\RR))\otimes\Cl_{q+7,3}\mapsto\sR_{q-3}^2\times\sC_{q+1}.
\end{equation}

\subsection{Pyritohedral Symmetry \(T_h\)}
The symmetry group is \({G=A_4\times\ZZ_2}\) with generators \(\hat{c}_3,\hat{c}_2,\hat{I}\) such that
\begin{equation}
\hat c_3^3=-1,\myspace \hat c_2^2=-1,\myspace (\hat{c}_2\hat c_3)^3=-1,\myspace \hat{I}^2=1.
\end{equation}
It acts on the 3D space as
\begin{equation}
\hat c_3^{\phantom{|}}\gamma_i\hat c_3^{-1}=\gamma_{i+1},\myspace \hat{c}_2^{\phantom{|}}\gamma_{1,2} \hat{c}_2^{-1}=-\gamma_{1,2},\myspace \hat I\gamma_i \hat I^{-1}=-\gamma_i.
\end{equation}
We set
\begin{equation}
c_3 = \tfrac{1+\gamma_1\gamma_2+\gamma_2\gamma_3+\gamma_3\gamma_1}{2}c_3,\myspace c_2 = -\hat{c}_2\gamma_1\gamma_2,\myspace I = \hat I \gamma_1\gamma_2\gamma_3,
\end{equation}
which satisfy
\begin{equation}
\begin{gathered}
c_3^{\phantom{|}} y_i c_3^{-1}=y_i,\myspace c_2^{\phantom{|}} y_i c_2^{-1}=y_i,\myspace Iy_iI^{-1}=-y_i,\\ \myspace c_3^3=1,\myspace c_2^2=1,\myspace (c_2 c_3)^3=1,\myspace I^2=1.
\end{gathered}
\end{equation}
We add \(I\) as a generator and get \({\RR[A_4]\otimes(\Cl_{p,q}\hat{\otimes}\RR[\ZZ_2])}\) and hence
\begin{equation}
(\RR\oplus\CC\oplus M_3(\RR))\otimes\Cl_{q+7,3+1}\mapsto\sR_{q-4}^2\times\sC_{q}.
\end{equation}

\subsection{Full Tetrahedral Symmetry \(T_d\)}
The symmetry group is \({G=S_4}\) with generators \(\hat{c}_3,\hat{s}_4\) such that
\begin{equation}
\hat c_3^3=-1,\myspace \hat s_4^4=-1,\myspace (\hat{s}_4\hat c_3)^2=-1.
\end{equation}
It acts on the 3D space as
\begin{equation}
\hat c_3^{\phantom{|}}\gamma_i\hat c_3^{-1}=\gamma_{i+1},\myspace \hat s_4^{\phantom{|}}\Big(\begin{smallmatrix}\gamma_1\\\gamma_2\\\gamma_3\end{smallmatrix}\Big)\hat s_4^{-1}=\Big(\begin{smallmatrix}-\gamma_2\\+\gamma_1\\-\gamma_3\end{smallmatrix}\Big).
\end{equation}
We set
\begin{equation}
c_3 = \tfrac{1+\gamma_1\gamma_2+\gamma_2\gamma_3+\gamma_3\gamma_1}{2}c_3,\myspace s_4 = e^{-\gamma_1\gamma_2\frac{\pi}{4}}\hat{s}_4\gamma_3,
\end{equation}
which satisfy
\begin{equation}
\begin{gathered}
c_3^{\phantom{|}} y_i c_3^{-1}=y_i,\myspace s_4^{\phantom{|}} y_i s_4^{-1}=-y_i,\\ c_3^3=1,\myspace s_4^4=1,\myspace (s_4 c_3)^2=1.
\end{gathered}
\end{equation}
Here we used the equality
\begin{equation}
\begin{aligned}
&(s_4 c_3)^2 \\&= -\tfrac{1-\gamma_1\gamma_2-\gamma_2\gamma_3-\gamma_3\gamma_1}{2}e^{-\gamma_1\gamma_2\frac{\pi}{4}} \hat s_4 \gamma_3 \hat c_3 e^{-\gamma_1\gamma_2\frac{\pi}{4}} \hat s_4 \gamma_3 \hat c_3 \\ &= -\tfrac{1-\gamma_1\gamma_2-\gamma_2\gamma_3-\gamma_3\gamma_1}{2}e^{-\gamma_2\gamma_3\frac{\pi}{4}}e^{-\gamma_1\gamma_2\frac{\pi}{4}}\gamma_2\gamma_3 \hat s_4 \hat c_3 \hat s_4 \hat c_3=1.
\end{aligned}
\end{equation}
We find that the ring of \({A_4\triangleleft S_4}\) commutes with \(\Cl_{p,q}\). Let us look at the algebra structure
\begin{equation}
\begin{matrix}
\RR[A_4]&=&\RR\oplus\CC\oplus M_3(\RR)\\
\hookdownarrow&&\hookdownarrow\\
\RR[S_4]&=&\RR^\ptwo\oplus M_2(\RR)\oplus M_3(\RR)^\ptwo\\
&&\veq\\
&&\Cl_{0,1}\oplus\Cl_{0,2}\oplus(M_3(\RR)\otimes\Cl_{0,1}),
\end{matrix}
\end{equation}
\begin{equation}
\begin{aligned}
\Cl_{p,q}\hat\otimes\RR[S_4]&=\Cl_{p,q}\hat\otimes(\Cl_{0,1}\oplus\Cl_{0,2}\oplus(M_3(\RR)\otimes\Cl_{0,1}))\\ &=((\RR\oplus M_3(\RR))\otimes\Cl_{p,q+1})\oplus\Cl_{p,q+2}.
\end{aligned}
\end{equation}
We thus get
\begin{multline}
((\RR\oplus M_3(\RR))\otimes\Cl_{q+7,3+1})\oplus\Cl_{q+7,3+2}\\
\mapsto\sR_{q-4}^2\times\sR_{q-5}.
\end{multline}

\subsection{Chiral Octahedral Symmetry \(O\)}
The symmetry group is \({G=S_4}\) with generators \(\hat{c}_3,\hat{c}_4\) such that
\begin{equation}
\hat c_3^3=-1,\myspace \hat c_4^4=-1,\myspace (\hat{c}_4\hat c_3)^2=-1.
\end{equation}
It acts on the 3D space as
\begin{equation}
\hat c_3^{\phantom{|}}\gamma_i\hat c_3^{-1}=\gamma_{i+1},\myspace \hat c_4^{\phantom{|}}(\begin{smallmatrix}\gamma_1\\\gamma_2\end{smallmatrix})\hat c_4^{-1}=(\begin{smallmatrix}+\gamma_2\\-\gamma_1\end{smallmatrix}).
\end{equation}
We set
\begin{equation}
c_3 = \tfrac{1+\gamma_1\gamma_2+\gamma_2\gamma_3+\gamma_3\gamma_1}{2}c_3,\myspace c_4 = e^{\gamma_1\gamma_2\frac{\pi}{4}}\hat{c}_4,
\end{equation}
which satisfy
\begin{equation}
\begin{gathered}
c_3^{\phantom{|}} y_i c_3^{-1}=y_i,\myspace c_4^{\phantom{|}} y_i c_4^{-1}=y_i,\\ c_3^3=1,\myspace c_4^4=1,\myspace (c_4 c_3)^2=1.
\end{gathered}
\end{equation}
Here we used the equality
\begin{equation}
\begin{aligned}
(c_4 c_3)^2 &= -\tfrac{1-\gamma_1\gamma_2-\gamma_2\gamma_3-\gamma_3\gamma_1}{2}e^{\gamma_1\gamma_2\frac{\pi}{4}} \hat c_4 \hat c_3 e^{\gamma_1\gamma_2\frac{\pi}{4}} \hat c_4 \hat c_3 \\&= -\tfrac{1-\gamma_1\gamma_2-\gamma_2\gamma_3-\gamma_3\gamma_1}{2}e^{\gamma_2\gamma_3\frac{\pi}{4}}e^{\gamma_1\gamma_2\frac{\pi}{4}} \hat c_4 \hat c_3 \hat c_4 \hat c_3=1.
\end{aligned}
\end{equation}
We thus get \({B=\RR[S_4]\otimes\Cl_{q+7,3}}\) and hence
\begin{equation}
(\RR^\ptwo\oplus M_2(\RR)\oplus M_3(\RR)^\ptwo)\otimes\Cl_{q+7,3}\mapsto\sR_{q-3}^5.
\end{equation}

\subsection{Full Octahedral Symmetry \(O_h\)}
The symmetry group is \({G=S_4\times\ZZ_2}\) with generators \(\hat{c}_3,\hat{c}_4,\hat I\) such that
\begin{equation}
\hat c_3^3=-1,\myspace \hat c_4^4=-1,\myspace (\hat{c}_4\hat c_3)^2=-1,\myspace \hat I^2=1.
\end{equation}
It acts on the 3D space as
\begin{equation}
\hat c_3^{\phantom{|}}\gamma_i\hat c_3^{-1}=\gamma_{i+1},\myspace \hat c_4^{\phantom{|}}(\begin{smallmatrix}\gamma_1\\\gamma_2\end{smallmatrix})\hat c_4^{-1}=(\begin{smallmatrix}+\gamma_2\\-\gamma_1\end{smallmatrix}),\myspace \hat I\gamma_i \hat I^{-1}=-\gamma_i.
\end{equation}
We set
\begin{equation}
c_3 = \tfrac{1+\gamma_1\gamma_2+\gamma_2\gamma_3+\gamma_3\gamma_1}{2}c_3,\myspace c_4 = e^{\gamma_1\gamma_2\frac{\pi}{4}}\hat{c}_4,\myspace I = \hat I \gamma_1\gamma_2\gamma_3,
\end{equation}
which satisfy
\begin{equation}
\begin{gathered}
c_3^{\phantom{|}} y_i c_3^{-1}=y_i,\myspace c_4^{\phantom{|}} y_i c_4^{-1}=y_i,\myspace Iy_iI^{-1}=-y_i,\\ c_3^3=1,\myspace c_4^4=1,\myspace (c_4 c_3)^2=1,\myspace I^2=1.
\end{gathered}
\end{equation}
We add \(I\) as a generator and get \({\RR[S_4]\otimes(\Cl_{p,q}\hat\otimes\RR[\ZZ_2])}\) and hence
\begin{equation}
(\RR^\ptwo\oplus M_2(\RR)\oplus M_3(\RR)^\ptwo)\otimes\Cl_{q+7,3+1}\mapsto\sR_{q-4}^5.
\end{equation}

\section{Symmorphic Layer Group Symmetry Classification}\label{app:layer}
\renewcommand{\theequation}{E\arabic{equation}}
\setcounter{equation}{0}
\renewcommand{\thetable}{E-\Roman{table}}
\setcounter{table}{0}

\begin{table*}[t]
\begin{equation*}
\begin{array}{l||c|c||c|c|c||c|c|c||c|c|c|c}
\mathrm{Sch\ddot{o}n.}	& C_1 & C_i,S_2 & D_1 & C_{1v} & D_{1d} & C_2 & C_{1h} & C_{2h} & D_2 & C_{2v} & D_{1h} & D_{2h}
\\ \hline\hline
& \sC_{q}	& \sC_{q}^2 & \sC_{q+1} & \sC_{q+1} & \sC_{q} & \sC_{q}^2 & \sC_q^2 & \sC_q^4 & \sC_{q+1}^2	& \sC_{q+1}^2 & \sC_{q} & \sC_{q}^2
\\ \hline\hline
\mathrm{A}		& \ZZ	& \ZZ^2	& 0		& 0		& \ZZ	& \ZZ^2		& \ZZ^2	& \ZZ^4	& 0			& 0			& \ZZ	& \ZZ^2
\\ \hline							
\mathrm{AIII} 	& 0		& 0		& \ZZ	& \ZZ	& 0		& 0			& 0		& 0		& \ZZ^2		& \ZZ^2		& 0		& 0		
\\ \hline\hline
& \sR_{q-2}	& \sC_{q} & \sR_{q-3} & \sR_{q-3} & \sR_{q-4} & \sR_{q-2}^2 & \sC_q & \sC_q^2 & \sR_{q-3}^2	& \sR_{q-3}^2 & \sR_{q-2} & \sR_{q-2}^2
\\ \hline\hline
\mathrm{AI} 	& 0		& \ZZ	& 0		& 0		& \ZZ	& 0			& \ZZ	& \ZZ^2	& 0			& 0			& 0		& 0					
\\ \hline
\mathrm{BDI} 	& 0		& 0		& 0		& 0		& 0		& 0			& 0		& 0		& 0			& 0			& 0		& 0					
\\ \hline
\mathrm{D} 		& \ZZ	& \ZZ	& 0		& 0		& 0		& \ZZ^2		& \ZZ	& \ZZ^2	& 0			& 0			& \ZZ	& \ZZ^2				
\\ \hline
\mathrm{DIII} 	& \ZZ_2	& 0		& \ZZ	& \ZZ	& 0		& \ZZ_2^2	& 0		& 0		& \ZZ^2		& \ZZ^2		& \ZZ_2	& \ZZ_2^2	
\\ \hline
\mathrm{AII} 	& \ZZ_2	& \ZZ	& \ZZ_2	& \ZZ_2	& \ZZ	& \ZZ_2^2	& \ZZ	& \ZZ^2	& \ZZ_2^2	& \ZZ_2^2	& \ZZ_2	& \ZZ_2^2
\\ \hline
\mathrm{CII} 	& 0		& 0		& \ZZ_2	& \ZZ_2	& \ZZ_2	& 0			& 0		& 0		& \ZZ_2^2	& \ZZ_2^2	& 0		& 0		
\\ \hline
\mathrm{C} 		& \ZZ	& \ZZ	& 0		& 0		& \ZZ_2	& \ZZ^2		& \ZZ 	& \ZZ^2	& 0			& 0			& \ZZ	& \ZZ^2	
\\ \hline
\mathrm{CI} 	& 0		& 0		& \ZZ	& \ZZ	& 0		& 0			& 0		& 0		& \ZZ^2		& \ZZ^2		& 0		& 0		
\end{array}
\end{equation*}
\caption{Bulk topological invariants and classifying spaces for the triclinic symmorphic layer group symmetries \(C_1,C_i\), the monoclinic inclined symmorphic layer group symmetries \(D_1,C_{1v},D_{1d}\), the monoclinic orthogonal symmorphic layer group symmetries \(C_2,C_{1h},C_{2h}\), and the orthorhombic symmorphic layer group symmetries \(D_2,C_{2v},D_{1h},D_{2h}\).}\label{tab:pp1}
\end{table*}

\begin{table*}[t]
\begin{equation*}
\begin{array}{l||c||c|c|c|c|c|c|c}
\mathrm{Sch\ddot{o}n.}	& C_1 & C_4 & S_4 & C_{4h} & D_4 & C_{4v} & D_{2d} & D_{4h}
\\ \hline\hline
& \sC_{q}	& \sC_{q}^4 & \sC_{q}^4 &\sC_{q}^8 & \sC_{q+1}^2\times\sC_{q} & \sC_{q+1}^2\times\sC_{q} & \sC_{q+1}^2\times\sC_{q} & \sC_{q}^4
\\ \hline\hline
\mathrm{A}		& \ZZ	& \ZZ^4				& \ZZ^4				& \ZZ^8	& \ZZ					& \ZZ				& \ZZ				& \ZZ^4
\\ \hline							
\mathrm{AIII} 	& 0		& 0					& 0					& 0		& \ZZ^2				& \ZZ^2				& \ZZ^2				& 0
\\ \hline\hline
& \sR_{q-2}	& \sR_{q-2}^2\times\sC_{q} & \sR_{q-2}^2\times\sC_{q} &\sC_{q}^4 & \sR_{q-3}^2\times\sR_{q-4} & \sR_{q-3}^2\times\sR_{q-4} & \sR_{q-3}^2\times\sR_{q-4} & \sR_{q-2}^4
\\ \hline\hline
\mathrm{AI} 	& 0		& \ZZ				& \ZZ				& \ZZ^4	& \ZZ					& \ZZ				& \ZZ				& 0	
\\ \hline
\mathrm{BDI} 	& 0		& 0					& 0					& 0		& 0					& 0					& 0					& 0	
\\ \hline
\mathrm{D} 		& \ZZ	& \ZZ^3				& \ZZ^3				& \ZZ^4	& 0					& 0					& 0					& \ZZ^4
\\ \hline
\mathrm{DIII} 	& \ZZ_2	& \ZZ_2^2			& \ZZ_2^2			& 0		& \ZZ^2				& \ZZ^2				& \ZZ^2				& \ZZ_2^4
\\ \hline
\mathrm{AII} 	& \ZZ_2	& \ZZ_2^2\times\ZZ	& \ZZ_2^2\times\ZZ	& \ZZ^4	& \ZZ_2^2\times\ZZ	& \ZZ_2^2\times\ZZ	& \ZZ_2^2\times\ZZ	& \ZZ_2^4
\\ \hline
\mathrm{CII} 	& 0		& 0					& 0					& 0		& \ZZ_2^3				& \ZZ_2^3			& \ZZ_2^3			& 0
\\ \hline
\mathrm{C} 		& \ZZ	& \ZZ^3				& \ZZ^3				& \ZZ^4	& \ZZ_2				& \ZZ_2				& \ZZ_2				& \ZZ^4
\\ \hline
\mathrm{CI} 	& 0		& 0					& 0					& 0		& \ZZ^2				& \ZZ^2				& \ZZ^2				& 0
\end{array}
\end{equation*}
\caption{Bulk topological invariants and classifying spaces for the tetragonal crystal system symmorphic layer group symmetries \(C_4,S_4,C_{4h},D_4,C_{4v},D_{2d},D_{4h}\).}
\end{table*}

\begin{table*}[t]
\begin{equation*}
\begin{array}{l||c||c|c|c|c|c}
\mathrm{Sch\ddot{o}n.}	& C_1 & C_6 & C_{3i},S_6 & D_3 & C_{3v} & D_{3d}
\\ \hline\hline
& \sC_{q}	& \sC_{q}^3 & \sC_{q}^6 & \sC_{q+1}\times\sC_{q} & \sC_{q+1}\times\sC_{q} & \sC_q^3
\\ \hline\hline
\mathrm{A}		& \ZZ	& \ZZ^3				& \ZZ^6	& \ZZ				& \ZZ					& \ZZ^3
\\ \hline							
\mathrm{AIII} 	& 0		& 0					& 0		& \ZZ				& \ZZ					& 0
\\ \hline\hline
& \sR_{q-2}	& \sR_{q-2}\times\sC_{q} & \sC_{q}^3 & \sR_{q-3}\times\sR_{q-4} & \sR_{q-3}\times\sR_{q-4} & \sR_{q-4}\times\sC_q
\\ \hline\hline
\mathrm{AI} 	& 0		& \ZZ				& \ZZ^3	& \ZZ				& \ZZ					& \ZZ^2
\\ \hline
\mathrm{BDI} 	& 0		& 0					& 0		& 0					& 0					& 0
\\ \hline
\mathrm{D} 		& \ZZ	& \ZZ^2				& \ZZ^3	& 0					& 0					& \ZZ
\\ \hline
\mathrm{DIII} 	& \ZZ_2	& \ZZ_2				& 0		& \ZZ				& \ZZ					& 0
\\ \hline
\mathrm{AII} 	& \ZZ_2	& \ZZ_2\times\ZZ	& \ZZ^3	& \ZZ_2\times\ZZ	& \ZZ_2\times\ZZ		& \ZZ^2
\\ \hline
\mathrm{CII} 	& 0		& 0					& 0		& \ZZ_2^2			& \ZZ_2^2				& \ZZ_2
\\ \hline
\mathrm{C} 		& \ZZ	& \ZZ^2				& \ZZ^3	& \ZZ_2				& \ZZ_2				& \ZZ_2\times\ZZ
\\ \hline
\mathrm{CI} 	& 0		& 0					& 0		& \ZZ				& \ZZ					& 0
\end{array}
\end{equation*}
\caption{Bulk topological invariants and classifying spaces for the trigonal crystal system symmorphic layer group symmetries \(C_3,C_{3i},D_3,C_{3v},D_{3d}\).}
\end{table*}

\begin{table*}[t]
\begin{equation*}
\begin{array}{l||c||c|c|c|c|c|c|c}
\mathrm{Sch\ddot{o}n.}	& C_1 & C_6 & C_{3h} & C_{6h} & D_6 & C_{6v} & D_{3h} & D_{6h}
\\ \hline\hline
& \sC_{q}	& \sC_{q}^6 & \sC_{q}^6 &\sC_{q}^{12} & \sC_{q+1}^2\times\sC_{q}^2 & \sC_{q+1}^2\times\sC_{q}^2 & \sC_{q}^3 & \sC_{q}^6
\\ \hline\hline
\mathrm{A}		& \ZZ	& \ZZ^6				& \ZZ^6	& \ZZ^{12}	& \ZZ^2				& \ZZ^2				& \ZZ^3		& \ZZ^6
\\ \hline							
\mathrm{AIII} 	& 0		& 0					& 0		& 0			& \ZZ^2				& \ZZ^2				& 0			& 0
\\ \hline\hline
& \sR_{q-2}	& \sR_{q-2}^2\times\sC_{q}^2 & \sC_{q}^3 &\sC_{q}^6 & \sR_{q-3}^2\times\sR_{q-4}^2 & \sR_{q-3}^2\times\sR_{q-4}^2 & \sR_{q-2}^3 & \sR_{q-2}^6
\\ \hline\hline
\mathrm{AI} 	& 0		& \ZZ^2				& \ZZ^3	& \ZZ^6		& \ZZ^2				& \ZZ^2				& 0			& 0	
\\ \hline
\mathrm{BDI} 	& 0		& 0					& 0		& 0			& 0					& 0					& 0			& 0	
\\ \hline
\mathrm{D} 		& \ZZ	& \ZZ^4				& \ZZ^3	& \ZZ^6		& 0					& 0					& \ZZ^3		& \ZZ^6
\\ \hline
\mathrm{DIII} 	& \ZZ_2	& \ZZ_2^2			& 0		& 0			& \ZZ^2				& \ZZ^2				& \ZZ_2^3	&\ZZ_2^6
\\ \hline
\mathrm{AII} 	& \ZZ_2	&\ZZ_2^2\times\ZZ^2	& \ZZ^3	& \ZZ^6		& \ZZ_2^2\times\ZZ^2	&\ZZ_2^2\times\ZZ^2	& \ZZ_2^3	&\ZZ_2^6
\\ \hline
\mathrm{CII} 	& 0		& 0					& 0		& 0			& \ZZ_2^4				& \ZZ_2^4			& 0			& 0
\\ \hline
\mathrm{C} 		& \ZZ	& \ZZ^4				& \ZZ^3	& \ZZ^6		& \ZZ_2^2				& \ZZ_2^2			& \ZZ^3		& \ZZ^6
\\ \hline
\mathrm{CI} 	& 0		& 0					& 0		& 0			& \ZZ^2				& \ZZ^2				& 0			& 0
\end{array}
\end{equation*}
\caption{Bulk topological invariants and classifying spaces for the hexagonal crystal system symmorphic layer group symmetries \(C_6,C_{3h},C_{6h},D_6,C_{6v},D_{3h},D_{6h}\).}
\label{tab:mm1}
\end{table*}

In this appendix we briefly present the topological bulk invariants of the symmorphic layer groups 
using analogous derivation to the 3D point groups presented in Sec.~\ref{sec:exam} and Appendix~\ref{app:pgf}.

Bulk topological invariants for all Altland-Zirnbauer symmetry classes are presented in Table~\ref{tab:pp1}~through~Table~\ref{tab:mm1}. All classifying spaces for all symmorphic layer group symmetries in all crystal systems are compactly presented in Table~\ref{tab:class}.

The layer groups are given by omitting the \(z\)-direction \(\gamma_3\) generator. Notice, that the equivalences of the 3D monoclinic and orthorhombic point groups no longer hold, i.e., \(C_2\neq D_1,~C_{1h}\neq C_{1v},~C_{2h}\neq D_{1d},~C_{2v}\neq D_{1h}\).

\subsubsection{Rotational Symmetry \(C_n\)}
This has no effect on the \(z\) direction and is hence just shifted by 1 due to the lack of the \(\gamma_3\) generator. We thus find
\begin{equation}
\sR_{q-2}^{r(n)}\times\sC_{q}^{c(n)}.
\end{equation}

\subsubsection{Rotoreflection Symmetry \(S_{2n}\)}
One has to modify the \(s\) generator to exclude \(\gamma_3\) by \(s'= e^{\gamma_1\gamma_2\frac{2\pi}{4n}}\hat s\). This however satisfies \(s' y_i s'^{-1}= y_i\) and \(s'^{2n}=(-1)^n\).

For \(n=1\) this is just a complex structure \(\CC=\RR\langle\sigma_h'\rangle\) and we get \(B=\Cl_{p+1,q}\otimes\CC\) and hence
\begin{equation}
\sC_{q}.
\end{equation}

For \(n=2\) we get \(B=\Cl_{p+1,q}\otimes\RR[\ZZ_4]\) and hence
\begin{equation}
\sR_{q-2}^2\times\sC_{q}.
\end{equation}

For \(n=3\) we have \(\RR\langle s'\rangle=\RR\langle -s'^2\rangle\otimes\RR\langle s'^3\rangle=\CC^{\oplus 3}\) and we get \(B=\Cl_{p+1,q}\otimes\CC^{\oplus 3}\) and hence
\begin{equation}
\sC_{q}^3.
\end{equation}

\subsubsection{Dipyramidal Symmetry \(C_{nh}\)}
One has to modify the \(\sigma_h\) generator to exclude \(\gamma_3\) by \(\sigma_h'=\hat{\sigma}_h\). This however satisfies \(\sigma_h'^2=-1\) and \(\sigma_h'y_i\sigma_h'^{-1}=y_i\). It thus adds a complex structure \(\CC=\RR\langle\sigma_h'\rangle\) to \(C_{n}\) and we get \(B=\Cl_{p+1,q}\otimes\RR[\ZZ_n]\otimes\CC\) and hence
\begin{equation}
\sC_{q}^{r(n)+2c(n)}.
\end{equation}

\subsubsection{Pyramidal Symmetry \(C_{nv}\)}
This has no effect on the \(z\) direction and is hence just shifted by 1 due to the lack of the \(\gamma_3\) generator. We thus find
\begin{equation}
\sR_{q-3}^{r(n)}\times\sR_{q-4}^{c(n)}.
\end{equation}

\subsubsection{Dihedral Symmetry \(D_n\)}
One has to modify the \(c_2\) generator to exclude \(\gamma_3\) by \(c_2'=\hat{c_2}\gamma_2\). This, however, satisfies \(c_2'^2=1\), \(c_2'y_ic_2'^{-1}=y_i\), and \((c_n c'_2)^2=1\) which makes it equivalent to \(C_{nv}\) by \(c_2'\leftrightarrow\sigma_v\) and hence we find
\begin{equation}
\sR_{q-3}^{r(n)}\times\sR_{q-4}^{c(n)}.
\end{equation}

\subsubsection{Antiprismatic Symmetry \(D_{nd}\)}
One has to modify the \(s\) generator to exclude \(\gamma_3\) by \(s'= e^{\gamma_1\gamma_2\frac{2\pi}{4n}}\hat s\). This, however, satisfies \(s' y_i s'^{-1}= y_i\) and \(s'^{2n}=(-1)^n\) as well as \((s'\sigma_v)^2=1\).

For \(n=1\) and we get \(B=\Cl_{p+1,q}\hat\otimes\Cl_{0,2}\) and hence
\begin{equation}
\sR_{q-4}.
\end{equation}

For \(n=2\) it is equivalent to \(C_{4v}\) by \(s'\leftrightarrow c\) and hence we find
\begin{equation}
\sR_{q-3}^2\times\sR_{q-4}.
\end{equation}

For \(n=3\) we have \(\RR\langle s',\sigma_v\rangle=\RR\langle -s'^2\rangle\otimes\RR\langle s'^3,\sigma_v\rangle\) and we get \(B=(\Cl_{p+1,q}\hat\otimes\Cl_{0,2})\otimes(\RR\oplus\CC)\) and hence
\begin{equation}
\sR_{q-4}\times\sC_q.
\end{equation}

\subsubsection{Prismatic Symmetry \(D_{nh}\)}
One has to modify both the \(c_2\) and the \(\sigma_h\) generators to exclude \(\gamma_3\) by \(c_2'=\hat{c_2}\gamma_2\) and \(\sigma_h'=\hat{\sigma}_h\). These, however, satisfy \(c_2'^2=1\), \(\sigma_h'^2=-1\) and \(c_2'y_ic_2'^{-1}=y_i\), \(\sigma_h'y_i\sigma_h'^{-1}=y_i\) as well as \((c_n c'_2)^2=1\), \((c'_2\sigma'_h)^2=1\). We have \(\RR\langle c_n,c_2',\sigma_h'\rangle=\RR\langle\sigma_h'\rangle\hat{\otimes}\RR\langle c_n,c_2'\rangle\), where \(\RR\langle\sigma_h'\rangle=\CC=\Cl_{1,0}\) and \(\RR\langle c_n,c_2'\rangle=\RR[\Dih_n]=\Cl_{0,1}^{\oplus r(n)}\oplus\Cl_{0,2}^{\oplus c(n)}\) is graded by \(\ZZ_2=\Dih_n/\ZZ_n\). We hence get \(\Cl_{p,q}\otimes(\Cl_{1,1}^{\oplus r(n)}\oplus\Cl_{1,2}^{\oplus c(n)})=\Cl_{p,q}\otimes M_2(\RR)^{\oplus r(n)+2c(n)}\) and find
\begin{equation}
\sR_{q-2}^{r(n)+2c(n)}.
\end{equation}

\section{Magnetic Point Groups - an Example}\label{app:mag}
\renewcommand{\theequation}{F\arabic{equation}}
\setcounter{equation}{0}
In this appendix, we demonstrate the generalizability of our method for the treatment of crystalline topological insulators and superconductors with magnetic point group symmetries.
 
Let us look at \(C_{2n}'\) with generators \({\hat{c}'=\hat{c}T}\) such that \(T^2=\epsilon_T\) and
\begin{equation}
\hat{c}'^{2n}=-\epsilon_T^n.
\end{equation}
It acts on the 3D space as
\begin{equation}
\hat c'(\begin{smallmatrix}\gamma_1\\\gamma_2\end{smallmatrix})\hat c'^{-1}=-(\begin{smallmatrix}\cos\frac{2\pi}{2n}&\sin\frac{2\pi}{2n}\\-\sin\frac{2\pi}{2n}&\cos\frac{2\pi}{2n}\end{smallmatrix})(\begin{smallmatrix}\gamma_1\\\gamma_2\end{smallmatrix}).
\end{equation}
We set
\begin{equation}
c'= e^{\gamma_1\gamma_2\frac{2\pi}{2n}}\hat{c}',
\end{equation}
which satisfies
\begin{equation}
c' y_i c'^{-1}=-y_i,\myspace c'^{2n}=\epsilon_T^n.
\end{equation}
The presence of charge conservation implies the existence of an ``imaginary" generator \({J^2=-1}\) such that \({c'Jc'^{-1}=-J}\). We thus get \({B=\Cl_{1,d}\hat{\otimes}\RR\langle c',J\rangle}\).

For \({n=1}\) we have \({\RR\langle c',J\rangle=\RR\langle c',c'J\rangle}\) and find
\begin{equation}
\begin{cases}
\Cl_{1,d}\hat{\otimes}\Cl_{0,2}\mapsto\sR_{-d}, & T^2=+1,\\
\Cl_{1,d}\hat{\otimes}\Cl_{2,0}\mapsto\sR_{4-d}, & T^2=-1.\\
\end{cases}
\end{equation}

For \({n=2}\) we split our algebra by the central element \(c'^2\) and find
\begin{equation}
\Cl_{1,d}\hat{\otimes}(\Cl_{0,2}\oplus\Cl_{2,0})\mapsto\sR_{-d}\times\sR_{4-d}.
\end{equation}

For \({n=3}\) we have
\({B=\RR\langle c'^2\rangle{\otimes}(\Cl_{1,d}\hat{\otimes}\RR\langle c'^3,c'^3J\rangle)}\)
and thus find
\begin{equation}
\begin{cases}
(\RR\oplus\CC){\otimes}(\Cl_{1,d}\hat{\otimes}\Cl_{0,2})\mapsto\sR_{-d}\times\sC_{d}, & T^2=+1,\\
(\RR\oplus\CC){\otimes}(\Cl_{1,d}\hat{\otimes}\Cl_{2,0})\mapsto\sR_{4-d}\times\sC_{d}, & T^2=-1.\\
\end{cases}
\end{equation}

As expected~\cite{kennedy2016bott}, in all cases, one may switch between \({\epsilon_T=\pm1}\) by tensoring with $\HH$.

\bibliographystyle{apsrev4-1}
%

\end{document}